\documentclass[11pt]{article}  
\usepackage{amssymb,graphicx} 
\usepackage {apalike}
\usepackage{undertilde}
\usepackage[normalem]{ulem} 
\usepackage{amscd}
\usepackage {amsfonts}
\usepackage{cancel}
\begin{document}

\title{Einstein's Physical Strategy, Energy Conservation, Symmetries, and Stability: ``but Grossmann \& I believed that the conservation laws were not satisfied''} 


\maketitle

\begin{center}
\author{J. Brian Pitts \\ Faculty of Philosophy, University of Cambridge 
\\ jbp25@cam.ac.uk }
\end{center}


\begin{abstract}

Recent work on the history of General Relativity by Renn, Sauer, Janssen \emph{et al}. shows that Einstein found his field equations partly by a physical strategy including the Newtonian limit, the electromagnetic analogy, and energy conservation.  Such themes are similar to those later used by particle physicists.  How do Einstein's physical strategy and the particle physics derivations compare?  What energy-momentum complex(es) did he use and why?  Did Einstein tie conservation to symmetries, and if so, to which?  How did his work relate to emerging knowledge (1911-14) of the canonical energy-momentum tensor and its translation-induced conservation?  After initially using energy-momentum tensors hand-crafted from the gravitational field equations, Einstein used an identity from his assumed linear coordinate covariance $x^{\mu^{\prime}}= M^{\mu}_{\nu} x^{\nu} $ to relate it to the canonical tensor.   Usually he avoided using matter Euler-Lagrange equations and so was not well positioned to use or reinvent the Herglotz-Mie-Born  understanding that the canonical tensor was conserved due to translation symmetries, a result with roots in Lagrange, Hamilton and Jacobi.   Whereas Mie and Born were concerned about the canonical tensor's asymmetry, Einstein did not need to worry because his \emph{Entwurf} Lagrangian is modeled not so much on Maxwell's theory (which avoids negative-energies but gets an asymmetric canonical tensor as a result) as on a scalar theory (the Newtonian limit). Einstein's theory thus has a symmetric canonical energy-momentum tensor.  But as a result, it also has 3 negative-energy field degrees of freedom (later called ``ghosts'' in particle physics).  Thus the \emph{Entwurf} theory fails a 1920s-30s \emph{a priori} particle physics stability test with antecedents in Lagrange's and Dirichlet's stability work; one might anticipate possible gravitational instability.

This critique of the \emph{Entwurf} theory can be compared with Einstein's 1915 critique of his \emph{Entwurf} theory for not admitting rotating coordinates and not getting Mercury's perihelion right.  One can live with absolute rotation but cannot live with instability. 

Particle physics also can be useful in the historiography of gravity and space-time, both in assessing the growth of objective knowledge and in suggesting novel lines of inquiry  to see whether and how Einstein faced the substantially mathematical issues later encountered in particle physics.  This topic can be a useful case study in the history of science on recently reconsidered questions of presentism, whiggism and the like.  Future work will show how the history of General Relativity, especially Noether's work, sheds light on particle physics.

\end{abstract}  %

keywords: conservation laws, uniformity of nature, General Relativity, Noether's theorems, energy-momentum tensor, Max Born

\tableofcontents

\section{Introduction}

\begin{quote} \ldots but Grossmann \& I believed that the conservation laws were not satisfied and  Newton's law did not result in first-order approximation. \cite[p. 160]{EinsteinBesso} \end{quote}  
 So wrote Einstein to Besso on December 10, 1915, explaining why the newly accepted `right' field equations had previously been rejected.  This first concern of Einstein and Grossmann, that the conservation laws were not satisfied, is actually rather mysterious, as will appear shortly.

General Relativity is traditionally credited to Einstein's mathematical strategy of Principles (equivalence, generalized relativity, general covariance, eventually Mach) in the contexts of  discovery and justification.  According to this common (often implicit) view, not only did Einstein discover his late 1915 field equations (note the success term ``discover'') using these principles, but also we today should believe his theory on the same grounds.  At any rate we today should have a reasonably high prior degree of belief in it on those grounds, and a low prior for theories violating those principles, so that empirical confirmation that has arisen since the mid-1910s can leave us rationally with a high degree of belief in his theory.  In an important sense the physics of space-time and gravity ``arrived'' in 1915-19, when the final field equations were found and striking empirical confirmation arrived in the form of the bending of light. Since that time little important has happened relevant to space-time philosophy, so understanding the 1910s makes one up to date.  
 This story (or perhaps caricature), though widely circulated partly due to Einstein's media penetration (reinforced on everything from T-shirts to the American Physical Society's robot-repellent process of selecting Einstein's face in order to read \emph{Physical Review} journals), can be challenged on both descriptive historical and normative philosophical grounds.

In fact a number of the leading historians of General Relativity have argued recently that Einstein found his field equations by a quite different route.  Indeed Einstein's ``physical strategy'' (it has been called, to contrast it with the more familiar ``mathematical strategy'' of Principles) has been recovered by historians  by a combination of meticulous study of his notebooks during the early-mid 1910s and taking seriously the published record of his work in that era.  Einstein's  physical strategy involves seeking relativistic gravitational field equations using the Newtonian limit, an analogy between gravitation and electromagnetism (the most developed relativistic field theory at the time), the coupling of all energy-momentum \emph{including gravity's} as a source for gravity, and energy-momentum conservation  \cite{NortonField,RennSauer,Janssen,BradingConserve,RennDwarfEmergence,Renn,JanssenRenn,RennSauerPathways}.  Thus the descriptive historical claim of the role of the mathematical strategy of principles in Einstein's process of discovery has been significantly exaggerated.  Indeed at least one  reason why has been offered:  Einstein re-wrote his own history in order to justify his decreasingly appreciated unified field theory quest \cite{vanDongenBook}.  Einstein might well also simply have come to believe that his mathematical strategy made the crucial difference in 1915.

Normative issues also  arise regarding the relative strengths of the physical and mathematical strategies as arguments.  
On the one hand, the mathematical strategy of Principles is  sometimes  seen as not  compelling \cite{NortonEliminative,StachelEliminative}. Thus one might look for something more compelling, such as eliminative inductions, Norton argued.  
On the other hand, Einstein's physical strategy also has its puzzles.  Just how Einstein arrived at (what we take to be) his definitive field equations is tortuous and paradoxical. Renn and Sauer find 
\begin{quote} what might be called the three epistemic paradoxes raised by the genesis of general relativity: \\
\hspace{.5in} \emph{The paradox of missing knowledge}\ldots.\\
\hspace{.5in} \emph{The paradox of deceitful heuristics}\ldots\\
\hspace{.5in} \emph{The paradox of discontinuous progress}.  How could general relativity with its nonclassical
consequences---such as the dependence of space and time on physical
interactions---be the outcome of classical and special-relativistic physics although
such features are incompatible with their conceptual frameworks?  \cite[pp. 118, 119]{RennSauerPathways} \end{quote} 
If the physical strategy and the mathematical strategy both have their limitations as arguments, where does that leave us?  

I suggest that we should not be afraid to find both strategies' lines of argument quite imperfect during the 1910s.  A great deal has been learned in the last 100 years, some of it empirical, some of it theoretical, which makes our epistemic situation very different from Einstein's in late 1915.  It also isn't clear that the most reasonable views to hold in 1915 were Einstein's.  Nordstr\"{o}m's (second) theory, for example, was simpler and hence arguably more plausible \emph{a priori}, before or after the Einstein-Fokker geometrization \cite{EinsteinFokker}.  While Einstein's late 1915 theory handled Mercury better than Nordstr\"{o}m's theory did, why must the evidence from Mercury more than offset the greater antecedent plausibility of Nordstr\"{o}m's theory by the standards of the day \cite{vonLaueNordstrom}? 
As von Laue put it,
\begin{quote} 
This agreement between two individual numbers [the perihelion prediction of Einstein and the Newcomb anomaly], achieved under conditions which cannot be arbitrarily altered, so that it seems uncertain whether the suppositions (specifically the assumption of two mass points) are fulfilled with sufficient accuracy, does not seem to be a sufficient reason, even though it is note-worthy, to change the whole physical conception of the world to the full extent as Einstein did in his theory. \cite[p. 182]{RoseveareMercury}%
\footnote{Diese \"{U}bereinstimmung zwischen zwei einzelnen
Zahlen, erhalten unter Bedingungen, an denen jede willk\"{u}rliche Ver\"{a}nderung
unm\"{o}glich ist, bei der uns selbst unsicher scheint, ob die Voraussetzungen
(wir denken an die Annahme zweier Massenpunkte) mit
ausreichender Genauigkeit erf\"{u}llt ist, scheint uns, so bemerkenswert
sie ist, doch kein hinreichender Grund, das gesamte physikalische
Weltbild von Grund aus zu \"{a}ndern, wie es die Einsteinsche Theorie
tut.  \cite[p. 269]{vonLaueNordstrom}}     
\end{quote} 
 In 1912-15 it wasn't so clear that a revised gravitational theory was needed anyway.  The matter-based zodiacal light hypothesis (or something close enough) had seemed to significant people (Newcomb's last view, expressed posthumously in 1912  \cite[pp. 226, 227]{Newcomb1912}, and Seeliger) to be doing well enough in addressing Mercury's problem \cite[pp. 69, 156]{RoseveareMercury}. Mercury's perihelion was not the only empirical difficulty for astronomy anyway \cite[p. 86]{RoseveareMercury}, so a gravitational solution for 
just that problem was not especially plausible in advance.  
 Zenker makes an interesting effort to relate this episode to formal philosophy of science \cite{ZenkerMinimal}.

 As Lakatos emphasized, if we want to speak of progress rather than mere change, we need normative standards \cite{LakatosHistory}.  If we distinguish our epistemic situation in 2015 from that of 1915 as we should, we both can and must find something(s) in the last 100 years to fill in the gap that 1915 cannot fill for us, in order to identify the rational progress made in space-time and gravitational theory.  Chang has emphasized the value of attending to contemporary knowledge in writing the history of science and integrating the history and philosophy of science \cite{ChangPhlogistonWhig,ChangWater}.


Fortunately there is at least one way to do that, a way largely unexplored, namely, using later particle physics arguments \cite{Kraichnan,Feynman}---which bring us to the normative challenge to the use of Einstein's principles.  Much of the reason that Einstein's work has generated a scholarly industry (as only a few scholars do), presumably, is the belief that Einstein's reasons for General Relativity, whatever they were, are also good reasons for us, indeed among the best that we have.   More broadly, General Relativity is one of the most impressive scientific theories, and it is at least plausible that something methodologically interesting can be learned from the process by which it arose.  But maybe our reasons are quite different from and even much better than Einstein's reasons.   Particle physicists know that Einstein's equations are what one naturally arrives for a local interacting massless spin-$2$ field, with good explanations for the detailed nonlinearity, general covariance, \emph{etc}. from nuts-and-bolts principles of (at least) Poincar\'{e}-covariant field theory. (Poincar\'{e} symmetry does not exclude a larger symmetry, as is especially clear from a Kleinian subtractive as opposed to Riemannian additive picture of geometry \cite{NortonKleinRiemann}.) 
 In 1939 it became possible to situate Einstein's theory \emph{vis-a-vis} the full range of relativistic wave equations and Lorentz group representations:  Pauli and Fierz recognized the equation for a massless spin 2 field as the source-free linearized Einstein equations \cite{Fierz,PauliFierz,FierzPauli}.  That same year Rosen wondered about deriving General Relativity's nonlinearities from an initially special relativistic starting point \cite{Rosen1}.  Work by Kraichnan, Gupta, Feynman, Weinberg, Deser \emph{et al.} eventually filled in the gaps, showing that, on pain of instability, Einstein's theory is basically the only option (eliminative induction, philosophers would say), with contributions by many authors \cite{Weyl,Papapetrou,Gupta,Kraichnan,Thirring,Halpern,HalpernComp,Feynman,Wyss,OP,Weinberg65,NSSexlField,Deser,VanN,DeserQG,SliBimGRG,BoulangerEsole}.  One could also consider \emph{massive} spin 2 gravity if it works \cite{PauliFierz,FierzPauli,OP,FMS}, an issue apparently settled negatively in 1970-72 but reopened recently and now very active \cite{HinterbichlerRMP,deRhamLRR}.\footnote{The inclusion of Weyl refers to what he actually showed, not to his intentions, which were to refute Birkhoff's flat space-time project.
 I thank Harvey Brown for discussion.} 
 To recall a punchy expression by Peter van Nieuwenhuizen, ``general relativity follows from special relativity by excluding ghosts''  \cite{VanN}, ghosts being negative-energy degrees of freedom.\footnote{Van Nieuwenhuizen's claim is slightly exaggerated---unimodular General Relativity is also an option, though it barely differs. So is something like a scalar-tensor theory with an arbitrary solution-dependent cosmological constant \cite{SliBimGRG}.  So is pure spin-$2$ massive gravity, it appeared in late 2010 \cite{deRhamGabadadze,HassanRosenNonlinear,HinterbichlerRMP,deRhamLRR}.  But in the early 1970s everyone read Boulware \& Deser \cite{TyutinMass,DeserMass} who said that linear pure spin-$2$ doesn't have the correct massless limit \cite{Zakharov,vDVmass1,vDVmass2,DeserMass,vanNmass}
and also  turns nonlinearly into spin-$2$ and spin-$0$, where the spin-$0$ is a ghost (negative energy, hence presumably unstable).  Apparently no one read Maheshwari \cite{MaheshwariIdentity}, who already exhibited one of the recent counterexamples that avoid the spin-$0$ ghost even nonlinearly \cite{MassiveGravity3}, so decades of opportunities were missed. Whether Vainshtein's conjectured \emph{nonperturbative} smooth massless limit \cite{Vainshtein,Vainshtein2} was believed is hard to say, but it didn't much matter as long as the nonlinear ghost issue seems insoluble as was claimed \cite{DeserMass}.  The bending of light has to be introduced as an empirical premise to refute Nordstr\"{o}m's theory, van Nieuwenhuizen rightly notes.}
According to Boulanger and Esole, 
\begin{quote} [i]t is well appreciated that general relativity is the unique way to consistently deform the
Pauli-Fierz action $\int \mathcal{L}_2$ for a free massless spin-2 field under the assumption of locality,
Poincar\'{e} invariance, preservation of the number of gauge symmetries and the number
of derivatives in $\mathcal{L}_2$ [references suppressed] \cite{BoulangerEsole}. \end{quote}

The point of invoking particle physics here, then, is to use it as a foil for Einstein's physical strategy, given that they share very similar core ideas, such as an electromagnetic analogy and crucial use of energy-momentum conservation. 
  Particle physics, one should recall,  was the mainstream tradition of scientific progress during the 1920s-50s, while ``General Relativity'' was a backwater topic in physics \cite{EisenstaedtEtiage,Eisenstaedt,SchutzGR}.  More specifically, it was a province for mathematicians and for speculative unified field theories that did not retain Einstein's late 1915 field equations \cite{vanDongenBook}.

As will appear below, Einstein's general ideas of using  the Newtonian limit, an electromagnetic analogy, and energy-momentum conservation were good ideas---in other hands they would in time yield a remarkably compelling argument leading almost uniquely to Einstein's field equations.  But Einstein's own development of them was a mixed enterprise, including surprising blunders reflecting ignorance not only of the literature of his day \cite{HerglotzConservationPoincare,Mie1,Mie3,BornEnergyMieHerglotz}, but also of results known already to Lagrange, Hamilton, Jacobi and others, to the effect that the conservation of energy is related to time translation invariance and the conservation of momentum to spatial translation invariance. 
According to Gorelik, ``There is no indication in Einstein's works that he was aware of the
connection between symmetry and conservation principles.'' \cite{GorelikConservation} Gorelik draws upon 1970s-80s Russian-language scholarship by Vladimir P. Vizgin \cite{Vizgin1972,Vizgin1975,Vizgin1981}.
If Einstein did not have knowledge on the topic, he nonetheless clearly had relevant (false) beliefs.  He appealed to the wrong symmetry (rigid general linear coordinate transformations), and secretly made use of the correct symmetry (rigid translation invariance), in deriving conservation laws.  There is also a severe disharmony, if not quite an inconsistency, in his occasional appeal to the equations of motion or field equations of matter.  He appeals to an action principle for a point particle, then generalizes to dust (to use our term) by analogy, then further generalizes by analogy a symmetric energy-momentum tensor for any sort of matter.  But he does not generalize the appeal to an action principle to continuous matter.  Why not?  Consequently Einstein set himself the extraordinary task of getting the gravitational field equations \emph{alone} to entail energy-momentum conservation.  This is not much better motivated that trying to get energy-momentum conservation for a system of two coupled oscillators while using the equations of motion of just one of them.  According to Kuhn, finding scientists' mistakes is a legitimate activity in the history of science (not discredited by Lakatosian whiggishness), and indeed a very illuminating activity \cite{LakatosKuhn}. One finds in Einstein's work both flashes of brilliance and dark patches.

This paper will sharpen the electromagnetic analogy by introducing an electromagnetic theory, ``\emph{Entwurf} electromagnetism,'' that is a much closer analog to Einstein's \emph{Entwurf} theory than is Maxwell's. Maxwell's theory entails the conservation of charge: take the 4-divergence, and watch the symmetric-antisymmetric cancellation on the left side because mixed partial derivatives are equal and $1-1=0$ \cite{DeserFlux}  But ``\emph{Entwurf} electromagnetism''  does not entail the conservation of charge, though of course it permits such conservation. (That isn't necessarily bad:  de Broglie-Proca massive electromagnetism \cite{DeserFlux,Jackson,UnderdeterminationPhoton} merely permits charge conservation, rather than requiring it; charged matter manages to conserve charge anyway using its own Euler-Lagrange equations.)  Relatedly, \emph{Entwurf} electromagnetism, unlike Maxwell or de Broglie-Proca, has various features and especially bugs which also have analogs in the \emph{Entwurf} theory, most notably  interacting positive- and negative-energy degrees of freedom.  Particle physics would later use such things (with the name ``ghosts'') as a more or less sufficient condition for instability, seeing such theories as doomed.  Excluding ``ghosts'' was found by Fierz and Pauli to be related to gauge freedom for massless particles/fields  at least for the kinetic (derivatives) part of the Lagrangian density \cite{Fierz,FierzPauli,Fierz2,Wentzel}.  Excluding ghosts implies (in the massless case) basically a divergenceless free field equation, because the divergence, if nonzero, would involve lower-spin degrees of freedom (such as a vector for a tensor theory, or a scalar for a vector theory) with negative energy. Divergenceless free field equations, in turn, necessitate that any source introduced must be conserved. The conservation of energy-momentum should hold only for gravity and everything else together, so gravity has to serve as part of its own source, implying nonlinearity of a quite specific sort.  In fact it is difficult to arrive at anything besides Einstein's equations as we know and love them (or maybe unimodular GR, scalar-tensor theories with an arbitrary cosmological constant, or pure spin-$2$ massive gravity, a topic much in the news since the late 2010s as a generalization of the Fierz-Pauli linear theory \cite{HinterbichlerRMP,deRhamLRR}).

This critique of the \emph{Entwurf} theory as doomed \emph{a priori} due to negative energies can be compared with Einstein's late 1915 self-critique of \emph{Entwurf} for not admitting rotating coordinates and not getting Mercury's perihelion right \cite{EinsteinSommerfeld,JanssenNemesis}. Admitting rotating coordinates was one of Einstein's many interesting conjectural desiderata, some of which led nowhere, so it isn't much of a criticism of the theory except \emph{ad hominem} to Einstein.  Not getting Mercury's perihelion right isn't much of a criticism, either:  Newton's and Nordstr\"{o}m's theories didn't get Mercury right, either, and a ``dark matter'' solution (though not dark, just heavier than it looked, zodiacal light) was hardly ruled out \cite{RoseveareMercury,EarmanJanssenMercury}.  It's a bit like a losing lottery ticket:  one would be very happy with a winning ticket, but one should not expect to win.  Nordstr\"{o}m's theory fared somewhat worse than Newton's theory regarding Mercury, but neither Nordstr\"{o}m nor Einstein seemed to hold that fact against Nordstr\"{o}m's theory \cite[pp. 155, 156]{RoseveareMercury}.

 This response is very much in line, incidentally, with Laudan's take on unsolved problems (problems that have not been solved by \emph{any} theory): 
\begin{quote} 
\emph{unsolved problems generally count as genuine problems only when they are no longer unsolved.} 
Until solved by some theory in a domain they are generally only ``potential'' problems rather than actual ones. [footnote suppressed] There are two factors chiefly responsible for this: \ldots Second, it is often the case that even when an effect has been well authenticated, \emph{it is very unclear to which domain of science it belongs} and, therefore, which theories should seek, or be expected, to solve it. \cite[pp. 18, 19, emphasis in the original]{LaudanProgress}\footnote{I thank the audience at EPSA in Duesseldorf for bringing Laudan's point to my attention.}   
\end{quote}


 \section{Conservation Quest and Symmetric Stress-Energy} 

Einstein wanted a local relativistic gravitational field theory that implied the conservation of energy-momentum.  Was gravity supposed to imply conservation by itself, or could matter help?  He seems not to have taken a consistent view on that crucial question.  Along the way he started using a variational formalism.  Thus he wrote down a Lagrangian density 
for his \emph{Entwurf} theory  \cite{EinsteinFormal} \cite[p. 869]{JanssenRenn}.  The \emph{Entwurf} theory  has special features that make energy-momentum easier to represent---namely, the canonical energy-momentum tensor is symmetric when the metric $g_{\mu\nu}$ is used to move an index. 
Relatedly, the \emph{Entwurf} also special vices, namely, negative-energy degrees of freedom.  Avoiding such things, one now knows, is why the canonical energy-momentum tensor for Maxwell's theory is asymmetric.  Whether such negative-energy degrees of freedom should have been seen as a vice in the mid-1910s is an interesting question.   Einstein seems to have little to say about the matter.   In any case one can make use of what we know now about classical field theory to interrogate the historical record to see how Einstein addressed issues that, we now know, were waiting for him in the mathematics.  
But it is getting a bit ahead to invoke variational principles and the canonical energy-momentum tensor initially, because Einstein did not use them initially.


A lengthy 1912 unpublished manuscript on special relativity, which has attracted some comment \cite{JanssenElectron,RoweLaue,LehmkuhlStressEnergy}, sheds useful light on the development of Einstein's thought on material stress-energy: 
\begin{quote} The general validity of the conservation laws and of the law of the inertia
of energy \ldots suggest that relations [deduced for electrodynamics] are to be ascribed a \emph{general} significance,
even though they were obtained in a very special case.
We owe this generalization, which is the most important new
advance in the theory of relativity, to the investigations of Minkowski,
Abraham, Planck, and Laue.[reference suppressed]  To every kind of material process we might study, we have to assign a symmetric tensor $(T_{\mu\nu})$\ldots. 
Each time, the problem to be solved consists in finding out how $(T_{\mu\nu})$ is to be
formed from the variables characterizing the processes studied. If several processes
that can be separated from one another as regards energy occur in the same space,
then we have to assign to each such individual process its own stress-energy tensor
$(T^{(1)}_{\mu\nu})$, etc., and to set $(T_{\mu\nu})$  equal to the sum of these individual tensors.
\cite[p. 79]{EinsteinSR1912} 
\end{quote} 
Electromagnetic (Minkowski) and continuum (Laue) stress-energy tensors were symmetric and not canonical   \cite{MinkowskiStress,LaueContinuum,NortonNordstrom}.  
Indeed the symmetric  stress-energy tensor that Einstein wanted was not essentially related to either of the two recipes now familiar to us, canonical or metric. 
At this stage Einstein also had covariant and contravariant stress-energy tensors for gravity itself \cite{EinsteinEntwurfGerman}.


\section{Intermittent Use of a Matter Lagrangian}

Einstein's occasional use of an action principle for matter is more mysterious than seems to have been noticed.  In his  1913 \emph{Entwurf} work he used a generally covariant particle matter action $-m \int dt \frac{ds}{dt}$, thus arriving at the geodesic equation, on the assumption that no  non-gravitational forces act.  He then quite appropriately and correctly generalized the law to what we now call dust: 
\begin{quote} In order to derive the law of motion of continuously distributed incoherent masses,
we calculate the momentum and the ponderomotive force per unit volume and apply
the law of the conservation of momentum.'' \cite{EinsteinEntwurfGerman} \end{quote} 
While ``balance law'' seems like a better choice of words than ``law of the conservation'' (and Einstein was often clearer on this point), his intention is clear.  Moving from particle to continuum by analogy,  $\Theta_{\mu\nu} = \rho_0 \frac{dx_{\mu}}{ds} \frac{dx_{\nu}}{ds}$. (He had not yet started writing contravariant indices ``up,'' though Grossmann mentions that notation of Ricci and Levi-Civita in a footnote.)      
\begin{quote} From the foregoing one surmises that the momentum-energy law will have the form (10)   
 $$\sum_{\mu\nu} \frac{\partial}{\partial x_{\nu} } (\sqrt{-g}g_{\sigma\mu} \Theta_{\mu\nu}) - \frac{1}{2} \sqrt{-g} \frac{\partial g_{\mu\nu} }{ \partial x_{\sigma} } \Theta_{\mu\nu}=0 . \hspace{.5in} (\sigma=1,2,3,4)$$ 
The first three of these equations ($\sigma=1,2,3$) express the momentum law, and the last one ($\sigma=4$) the energy law.  It turns out that these equations are in fact covariant with respect to arbitrary substitutions. [footnote suppressed] Also, the equations of motion of the material point from which we started out can be rederived from these equations by integrating over the thread of flow.
\cite{EinsteinEntwurfGerman}
\end{quote} 

It is worth noting something that the antiquated notation tends to hide, namely, that Einstein's conserved quantity is a weight $1$ tensor density with mixed indices, one ``up'' from $\Theta_{\mu\nu}$ (as required for a partial divergence) and one ``down'' from $g_{\sigma\mu}$.  This sort of quantity is as good as any and better than most if one is hoping to take a covariant derivative and have few Christoffel symbol terms (as one can see in Grossmann's mathematical part of the paper).  It also has the same transformation rule as the canonical energy-momentum tensor (mixed tensor density of weight $1$), which was in the process of being invented by Herglotz, Mie and Born \cite{HerglotzConservationPoincare,Mie1,Mie3,BornEnergyMieHerglotz}, apparently unbeknownst to Einstein.  So far, so good.

Einstein goes on: 
\begin{quote}  We call the tensor $\Theta_{\mu\nu}$ the (contravariant) \emph{stress-energy tensor of the material flow}.  We ascribe to equation (10) a validity range that goes far beyond the special case of the flow of incoherent masses.  The equation represents in general the energy balance between the gravitational field and an arbitrary material process; one has only to substitute for $\theta_{\mu\nu}$ the stress-energy tensor corresponding to the material system under consideration. The first sum in the equation contains the space derivatives of the stresses or of the density of the energy flow, and the time derivatives of the momentum density or of the energy density; the second sum is an expression for the effects exerted by the gravitational field on the material process. \cite{EinsteinEntwurfGerman} \end{quote}  
A bit later he listed ``mechanical, electrical, and other processes''.  Perhaps at this point one notices more clearly what Einstein did not do in introducing the incoherent mass flow, namely, \emph{introduce any Lagrangian density to give equations of motion/field equations for matter}.  
Thus Einstein supplied no relation between $\Theta^{\mu\nu}$ and a continuous matter action. One further notices that he didn't derive the particulate matter stress-energy tensor from the Lagrangian, either; the particle Lagrangian yielded equations of motion (the geodesic equation), but the material stress-energy tensor is not derived from it.  Rather, just the conservation/balance equations are derived using the matter equations of motion that follow from the Lagrangian.

His use of a matter action for a particle indicates that he didn't object to action principles; before long he would also adopt one for gravity.  Why, then, did he not use one for continuous matter?  
Did he perhaps think that an \emph{explicit} Lagrangian density was required, so that it would be useless merely to assume that one existed without writing it down?  Relativistic fluids are admittedly not simple even for more modern authors \cite{RayFluid,SchutzSorkin}, and some of the complication applies already for dust, though there was progress rather early \cite{HerglotzConservationPoincare,BornEnergyMieHerglotz,LorentzEnergy,NordstromGREnergy}. 
If this omission of a matter Lagrangian density is only somewhat mysterious in 1913 (when various ingredients are missing or novel), it will become much mysterious once Einstein relies more heavily on Lagrangian densities.  
The would-be commutative diagram, though wobbly on the particle side (left) due to putting in the stress-energy tensor definition by hand (``exogenous''), is outright broken on the continuum side (right) due to the empty top-right corner:    
$$\begin{CD}
Lagrangian\,for\,particle\,in\,gravity   @>{continuum}>by\,analogy??>   \,  [NOTHING] \\
@Vgeodesic\,equationV\,of\,motion V                                        @Vequation\,Vof\, motion??V\\
exogenous\,\Theta^{\mu\nu}   @>{continuum}>by\,analogy>  exogenous\,\Theta^{\mu\nu}  \\
satisfies\,balance\,law             @.                                satisfies\,balance\,law   
  	 \end{CD}$$
Over time the mathematical technology employed would be improved: notation with contravariant indices up (thus making many aspects of general covariance evident by inspection),  a gravitational Lagrangian density, and expression of the material stress-energy balance law using a covariant derivative, for example.

But what never changed while Einstein sought the field equations through late 1915 is the missing matter Lagrangian in the top right corner of this diagram. (A matter Lagrangian does appear in 1916 \cite{EinsteinHamiltonGerman}, although it isn't used much, as Einstein emphasizes.) That absence becomes important when one compares to Born's canonical tensor formalism.  Why?  Because it will become obvious that, using Born's formalism, conservation laws follow automatically from the gravitational \emph{and material} equations of motion assuming merely \emph{rigid space-time translation invariance}, that is, the mere \emph{uniformity of nature}, a rather humble and familiar set of symmetries known for most or all of a century to do such jobs (\cite{LagrangeMecaniqueAnalytiqueNouvelle} (translated as  \cite[pp. 180-183]{LagrangeEnglish} for energy, \cite{HamiltonConservation,Jacobi} for momentum).

  Einstein, by contrast, uses the wrong symmetry (rigid general linear invariance, a symmetry of his own devising) while not appealing to material equations of motion.  That he gets conservation laws at all will result only because he sneaks space-time translation invariance in without noticing! Below I will introduce a counterexample that is invariant under rigid general linear transformations but depends on the coordinates and so violates the conservation laws.  To get conservation, it isn't necessary to write out explicitly what the matter Lagrangian density is; it is sufficient that it exist (along with the gravitational Lagrangian, which Einstein will adopt) and be \emph{space-time translation invariant}, in order for conservation to follow.  By not making that move (which would have been evident by comparison to the emerging canonical tensor formalism \cite{HerglotzConservationPoincare,Mie1,Mie3,BornEnergyMieHerglotz}, Einstein had to work a good deal harder.  He falsely believed that conservation was tied to the detailed form of the gravitational field equations. In fact his criterion of satisfying energy-momentum conservation was nearly trivial, because he theory satisfying the uniformity of nature and having a variational formulation would satisfy it.

When Einstein and Grossmann 
 later tie their balance law to covariant differentiation and use a Gothic $\mathfrak{T}^{\nu}_{\sigma}$ for the material stress-energy tensor \emph{density}, the material energy balance law will be writable as $\nabla_{\nu} \sqrt{-g} \theta^{\nu}_{\sigma}=0,$ \emph{i.e.}, $\nabla_{\nu} \mathfrak{T}^{\nu}_{\sigma}=0.$ 
 Einstein is often careful---not always, but more often  than writers today are---to note that this is an equation of balance, not  conservation. (The point has been made at times, including \cite{Landau,GorelikConservation,OhanianEnergyNonminimal}.\footnote{Ohanian's use of a weight $0$ rather than weight $1$ expression is suboptimal, however, because it obscures the difference between a Christoffel term that cannot be removed and one that can.  Similar remarks hold for Gorelik's symmetric weight $2$ expression. To obtain a tensorial expression, one needs a covariant divergence.   To obtain a continuity equation that can be integrated into a conservation law, one needs a partial divergence.  These requirements are compatible for contravariant vector densities of weight $1$ (and their axial cousins), but otherwise conflict.  To turn a displacement (tangent) vector into a contravariant vector density of weight $1$, one can multiply by a mixed tensor density of weight $1$.  Not coincidentally, the canonical energy-momentum tensor is just that sort of entity, at least in simple enough field theories including special relativistic ones.  In GR one can try to do the same thing, but complications arise  \cite{GoldbergConservation,TrautmanConserve,SchutzSorkin}. })  Gravity affects matter, so one would expect conservation to be violated unless some way to keeping track of gravitational energy-momentum is used.


\section{\emph{Entwurf} Derivation of Gravitational Field Equations} 

The derivation of the gravitational field equations in the \emph{Entwurf} contains a fascinating mixture of good ideas that would bear fruit later in the particle physics tradition, and subtle interesting error.  Generalizing the Newtonian field equations
$\nabla^2 \phi = 4 \pi k \rho$ would presumably lead to  
\begin{quote} the form 
$$(11)  \hspace{.5in}  \kappa \cdot \Theta_{\mu\nu} = \Gamma_{\mu\nu} , $$
where $\kappa$ is a constant and $\Gamma_{\mu\nu}$ a second-rank contravariant tensor derived from the fundamental tensor $_{\mu\nu}$ by differential operations.  In line with the Newton-Poisson law one would be inclined to require that these equations (11) be \emph{second order}. \cite{EinsteinEntwurfGerman} \end{quote}  Such an equation is seemingly inconsistent with general covariance, he says; maybe there are fourth derivatives, but that seems premature to consider.  [Purely algebraic terms, as in the 1890s Seeliger-Neumann long-range modification of Einstein's equations and the first half of Einstein's 1917 cosmological constant paper \cite{EinsteinCosmologicalGerman,NortonWoes,ScalarGravityPhil}, are also not entertained.]   ``Besides, it should be emphasized that we have no basis whatsoever for assuming a general covariance of the gravitational equations. [footnote suppressed]'' \cite{EinsteinEntwurfGerman}  Instead tensoriality under arbitrary linear transformations is required.

Einstein wants to find the nonlinear part of the gravitational field equations to ensure the conservation of energy-momentum---a nontrivial task when one is not using a variational principle.  His method is clever and largely sound, but may contain an interesting and important mistake that will make energy-momentum conservation seem much more difficult to achieve than it actually is, as will appear.   (If Einstein does not make that mistake here, he will make it eventually; the roots of ambiguity might appear before the tree.)  
Recalling equation   (10), the material energy-momentum balance law, it follows that    
 $$\sum_{\mu\nu}  \frac{1}{2} \sqrt{-g} \frac{\partial g_{\mu\nu} }{ \partial x_{\sigma} } \Theta_{\mu\nu}$$ 
(the opposite of the second, non-divergence term) ``is the momentum (or energy) imparted by the gravitational field to the matter per unit volume.'' 
The first term, $\sum_{\mu\nu} \frac{\partial}{\partial x_{\nu} } (\sqrt{-g}g_{\sigma\mu} \Theta_{\mu\nu}) $ 
presumably seems already acceptable, being equated to a divergence.  It isn't \emph{identically} a divergence, I note,  but an expression involving the metric and matter fields and derivatives thereof, proportional to the matter equations of motion and thus $0$ only if one uses the matter equations of motion.  Does Einstein lose track of the role of the matter equations in getting here?  Resuming with Einstein:
\begin{quote}  For the energy-momentum law to be satisfied, the differential expressions $\Gamma_{\mu\nu}$ of the fundamental quantities $\gamma_{\mu\nu}$ [the inverse metric $g^{\mu\nu}$] that enter the gravitational equations 
$$ \kappa \Theta_{\mu\nu} = \Gamma_{\mu\nu}$$
must be chosen such that 
$$ \frac{1}{2 \kappa} \sum_{\mu\nu} \sqrt{-g} \cdot \frac{\partial g_{\mu\nu}}{\partial x_{\sigma} } \Gamma_{\mu\nu}$$
can be rewritten in such a way that it appears as the sum of differential quotients. [that is, as a divergence] \cite[p. 163]{EinsteinEntwurfGerman}  \end{quote}  
Einstein has dropped the first term (the coordinate divergence) from equation 10 (as if forgetting that it involves the matter equations of motion), substituted the gravitational field equations into the second term, and asked for the resulting modified second term to be a divergence.  Grossmann in the mathematical part supplies the mathematical manipulations.  If Einstein's goal is merely to ensure that the gravitational equations \emph{in the absence of matter} conserve energy-momentum, all is innocent.  If he wishes to achieve energy-momentum conservation for gravity + matter with the help of the matter equations of motion as well, again all is fine.  Eventually he uses the gravitational equations and equation 10 to get energy-momentum conservation for gravity + matter; hence much depends on how clearly one understands the roots of equation 10.  

   The lengthy equation 12 is an identity with the left side being a divergence and the right side (up to a constant factor) being a product of $\sqrt{-g} \frac{\partial g_{\mu\nu}}{\partial x_{\sigma} } $ and a term in large curly braces containing the sum/difference of four terms.  The third and fourth terms  in curly braces in equation 12 are isolated in equation 13 (which I divide by $2$ for convenience and express in more modern notation, including the summation convention) as  
$$ -\kappa \theta^{\mu\nu}  = \frac{1}{2} g^{\alpha\mu} g^{\beta\nu} \frac{\partial g_{\tau\rho}}{\partial x^{\alpha} } \frac{ \partial g^{\tau\rho} }{\partial x^{\beta} } - \frac{1}{4} g^{\mu\nu} g^{\alpha\beta} \frac{\partial g_{\tau\rho} }{\partial x^{\alpha}} \frac{ \partial g^{\tau\rho} }{\partial x^{\beta} }.$$ 
This $\theta^{\mu\nu}$ is designated the `` `\emph{contravariant stress-energy tensor of the gravitational field}.' '' (emphasis and quotation in the original)  A corresponding covariant tensor $t_{\mu\nu}$ is achieved by lowering both indices with $g_{\mu\nu}.$ 
Note that these entities are symmetric due to 
$$ \frac{ \partial g^{\tau\rho} }{\partial x^{\beta}} = -g^{\tau\sigma} g^{\lambda\rho} \frac{ \partial g_{\sigma\lambda} }{\partial x^{\beta}} , $$ and that symmetry requires having the indices in the same place, so they differ from the mixed tensor density expression earlier by moving an index with the position-dependent metric or inverse metric.  Thus these symmetric entities are not directly useful for conservation laws.  These entities are also tensors under general rigid linear (or, for that matter, affine) coordinate transformations, which is all the invariance to which Einstein and Grossmann presently aspire (though the translation part of affine symmetry is tacit).  
The first and second terms on the right side of identity 12 are collected together and named $\Delta_{\mu\nu}(\gamma),$ a sort of nonlinear generalization of the d'Alembertian; it appears to serve also as a receptacle for whatever part of the right side of equation 12 did not look like part of a stress-energy tensor.  Using the stress-energy tensor to re-express the left side of equation 12 also, one indeed gets
equation 12a, which in modernized notation is
$$  \frac{\partial}{\partial x^{\nu}}  \left( \sqrt{-g} g_{\sigma\mu} \kappa \theta^{\mu\nu} \right)= \frac{1}{2} \sqrt{-g} \frac{\partial g_{\mu\nu} }{\partial x^{\sigma}} (-\Delta^{\mu\nu} + \kappa \theta^{\mu\nu}).$$
Einstein perceives an interesting analogy between the role of the material stress-energy tensor $\Theta^{\mu\nu}$ in the ``conservation law (10) for matter'' (really a balance law) and  the role of the gravitational stress-energy tensor $\theta^{\mu\nu}$ in  ``conservation law (12a) for the gravitational field\ldots. [T]his is a noteworthy circumstance considering the difference in the derivation of the two laws.''  I confess that I don't see the point, unless one introduces stronger independently motivated restrictions on $\Delta^{\mu\nu}.$  Such independent conditions do indeed hold in the later particle physics tradition \cite{Kraichnan,Deser,SliBimGRG}, however, in which the corresponding entity is linear, hence describing a free (non-interacting) gravitational field rather than self-interacting one. Linearity is partly dependent on one's choice of field variables such as $g_{\mu\nu}$ or $g^{\mu\nu},$ so it is possible to have a free field that doesn't look free.  I see no reason to think that such a hidden virtue applies here, however.  

Using the identity result $\Gamma^{\mu\nu} = \Delta^{\mu\nu}   - \kappa \theta^{\mu\nu}$ (equation 17 in updated notation), 
the gravitational field equations (11) above $ \Gamma^{\mu\nu} =  \kappa  \Theta_{\mu\nu} $ (updated)
take the form (18 in updated notation) $$ \Delta^{\mu\nu} = \kappa(\Theta^{\mu\nu} + \theta^{\mu\nu}). $$
Einstein and Grossmann remark: 
\begin{quote} 
These equations satisfy a requirement that, in, our opinion, must be imposed on a relativity theory of gravitation; that is to say, they show that the tensor $\theta_{\mu\nu}$ of the gravitational field acts as a field generator in the same way as the tensor $\Theta^{\mu\nu}$ of the material processes. An exceptional position of gravitational energy in comparison with all other kinds of energies would lead to untenable consequences. 
\cite{EinsteinEntwurfGerman} \end{quote}
 This \emph{universal coupling} principle is a cornerstone of the particle physics derivations such as \cite{Kraichnan,Deser,SliBimGRG}.  But it looks as though the Einstein-Grossmann version is trivial, absent a better reason to regard the form of $\Delta^{\mu\nu}$ as independently plausible rather than merely as a waste receptacle for all terms that don't look like a stress-energy tensor.

Einstein now turns to recovering the energy-momentum conservation law in his melodramatic way that neglects the matter equations and the full significance of the uniformity of nature.
\begin{quote} 
Adding equations (10) and (12a)  while taking into account equation (18), one finds
$$ (19) \sum_{\mu\nu} \frac{\partial }{\partial x^{\nu} } { \sqrt{-g} \cdot g_{\sigma\mu} (\Theta_{\mu\nu} + \theta_{\mu\nu})}=0 \hspace{.5in} (\sigma=1,2,3,4)$$
\emph{This shows that the conservation laws hold for the matter and the gravitational field taken together.}
\end{quote}
Einstein's result here is both correct and satisfying. 
He also presents a covariant form with a notation that looks more familiar to modern readers in his equation (22), which in updated notation (and correction of the differentiation index from $\nu$) is   
$$ \frac{\partial}{\partial x^{\sigma}} (\sqrt{-g}g^{\sigma\mu} (T_{\mu\nu} + t_{\mu\nu}))=0.$$ 
Both conservation laws are somewhat indirect in terms of symmetric (and weight $0$) stress-energy tensors, because the conservation laws are most naturally written in mixed weight $1$ form, as becomes clear later with a variational principle.  
One recalls that coordinate divergences integrate into conserved quantities, whereas covariant divergences (except where those few special cases in which they are coordinate divergences, such as weight 1 contravariant vector densities \cite{Anderson}) do not \cite[p. 280]{Landau}. 
 Unfortunately it seems plausible here, and becomes clear later when he uses a variational principle, that he doesn't realize that he has worked far harder than necessary to achieve this result.  As Lagrange and others knew much earlier, conservation laws hold due to the spatio-temporal uniformity of nature and the equations of motion/field equations for \emph{everything that has energy and momentum}.  If one does not keep track of when one uses the material equations of motion/field equations, one might mistakenly think that they  are not used, leaving the gravitational field equations to do all the work.


\section{Partial Shift to Variational Principle}   

Einstein and Grossmann  make a partial shift to a variational principle in the paper ``Covariance Properties of the Field Equations of the Theory of Gravitation Based on the Generalized Theory of Relativity''  \cite{EinsteinCovariance}.  
To be specific, he introduces a Lagrangian density for  gravity \emph{but not matter}.  Contravariant indices are still written down, at the beginning of the paper, but soon are moved up apart from the coordinate differentials $dx_{\nu}.$  For convenience I will modernize the notation without further comment.  
The gravitational field equations come from varying the action (using $H$ as in ``Hamilton'' for the Lagrangian density---in this paper  $H$ has absorbed the factor of $\sqrt{-g}$)
$$ \int d\tau ( \delta H - 2 \chi \sqrt{-g} T_{\mu\nu} \delta g^{\mu\nu})=0.$$ 
It would be convenient if they could say that  $\sqrt{-g} T_{\mu\nu}$ was the derivative of the Lagrangian density with respect to $g^{\mu\nu},$ as Hilbert later would.  As matters stand, 
Einstein and Grossmann rather conspicuously avoid introducing a matter Lagrangian density and deriving matter field equations from it. ``In the Lagrangian, matter thus appears simply `black-boxed,' rather than as an expression explicitly involving some set of variables describing the constitution of matter\ldots.'' \cite[p. 859]{RennStachel}. 

Later in the year (October/November) in ``The Formal Foundation of the General Theory of Relativity,''  Einstein  draws on the Minkowski-Laue special relativistic material stress-energy, which is symmetric and includes anistropic pressure/stress \cite{EinsteinFormal}.
 Einstein, however, argues for the mixed, weight $1$ tensor density $\mathfrak{T}^{\nu}_{\sigma}$, which, one recalls, is not symmetric and is similar to what we now know as the canonical tensor  \cite{HerglotzConservationPoincare,Mie1,Mie3,BornEnergyMieHerglotz,NordstromGREnergy}.     

Einstein rightly insists on deriving an energy-momentum conservation law and posits a gravitational energy-momentum $\frak{t}^{\nu}_{\sigma}$ to complement the material stress-energy $\frak{T}^{\nu}_{\sigma}$ to help achieve it.
In the absence of external forces 
\begin{quote} 
one has to demand that for the material system and associated gravitational field combined, some theorems remain that express the
constancy of total momentum and total energy of matter plus gravitational field. This
leads to the statement that a complex of quantities $\frak{t}^{\nu}_{\sigma}$  must exist for the gravitational
field such that the equations
$$ \sum_{\nu} \frac{\partial ( \frak{T}^{\nu}_{\sigma} + \frak{t}^{\nu}_{\sigma})}{\partial x_{\nu} }  \hspace{.5in}          (42.c)$$
apply. This point can only be discussed in more detail when the differential laws of
the gravitational field have been established. \cite{EinsteinFormal} \end{quote} 
This last sentence is important, albeit in a worrisome way.  It suggests that Einstein thinks that the detailed gravitational field equations play a key role in achieving conservation, not merely a rather weak structural property (such as translation invariance, a.k.a. the uniformity of nature) as is actually the case.


\section{Translation Symmetry Smuggling} 

After further discussion of the conservation laws for several types of matter \cite{EinsteinFormal}, a version of the hole argument appears as an argument against general covariance for the gravitational field equations, notwithstanding the general covariance of the matter field equations.
Thus one is satisfied with ``linear''  transformations. The word ``linear'' is ambiguous between strictly linear (like $y=mx$) and affine (like $y=mx+b$) interpretations, which is interesting given Einstein's tendency to neglect the role of translations.
Later, in section 15, the invariance of the  action under strictly linear transformation properties is discussed, resulting in a condition  $2 S^{\nu}_{\sigma} \equiv 0$  for a scalar action under linear coordinate transformations.  Notationally, the Lagrangian density is now $\sqrt{-g} H$: $H$ itself has reverted to being a scalar. 
Using the abbreviation $g^{\mu\nu}_{\sigma}$ for $\frac{\partial g^{\mu\nu} }{\partial x_{\sigma} },$ one has equation 76a:
$$ S^{\nu}_{\sigma} = g^{\nu\tau} \frac{\partial (H \sqrt{-g}) }{\partial g^{\sigma\tau}} + g^{\nu\tau}_{\mu} \frac{\partial (H \sqrt{-g}) }{\partial g^{\sigma\tau}_{\mu} } +  \frac{1}{2} \delta^{\nu}_{\sigma} H \sqrt{-g} -\frac{1}{2} g^{\mu\tau}_{\sigma} \frac{ \partial (H\sqrt{-g}) }{\partial g^{\mu\tau}_{\nu} }  $$

This identity and other considerations involving preferred coordinates will be used to infer the conservation of energy-momentum.  This derivation of equation 76a, unfortunately, is a red herring.  It smuggles rigid translation symmetry without mentioning it, apart
from the restriction that it depend only on the inverse metric and its first partial derivative---which  excludes explicit dependence on the coordinates.    Thus Einstein is tacitly using affine, not strictly linear coordinate transformations.  Translations should hardly have been controversial, so extending the symmetry group from linear to affine is entirely reasonable.  Unfortunately Einstein seemed not to notice that he was doing so, and hence was confused about the relationship between symmetries and conservation laws.   Did he  think that \emph{more} symmetries ($16$ linear or $20$ affine \emph{vs.} $6$ Lorentz or $10$ Poincar\'{e}) could imply \emph{fewer} conservation laws? If not, why not rest in the knowledge that Poincar\'{e} invariance already ensures the conservation laws?  
 Clearly in light of Noether's first theorem \cite{Noether} that cannot be the correct view, whether or not it was Einstein's at the time.  While one cannot hold Einstein responsible for mathematics developed a few years later, the antiquity (first half of the 19th century, as well as recent continuum work \cite{HerglotzConservationPoincare}) of the connection between translation symmetries and conservation laws makes it difficult to hold him blameless here.  The larger affine group gives if anything \emph{more} conservation laws, not fewer, than a merely Poincar\'{e}-covariant theory would have.  At worst, a larger symmetry group could lead to  conservation laws that are trivial, redundant or weird---but still true. 
 Did he \emph{forget translations} and really use only \emph{linear} coordinate covariance $x^{\mu^{\prime}} = M^{\mu}_{\nu} x^{\nu}$ \emph{vs.} affine $x^{\mu^{\prime}} = M^{\mu}_{\nu} x^{\nu} + N^{\mu}$? ($M^{\mu}_{\nu}$ and $ N^{\mu}$ constant)     %

Rigid translations relative to  one coordinate system  are rigid translations relative to any affinely related coordinate system as well.  Hence assuming translation invariance in one coordinate system implies translation invariance in all of the desired coordinate systems.  But given that translation invariance implies energy-momentum conservation \emph{via} the (now) well known derivation of the canonical energy-momentum tensor \cite{Goldstein}, 
 the specifically linear coordinate symmetry of the Lagrangian density must be doing something else.

To see where translations are smuggled in, one can work through the derivation.   Of the Lagrangian density he writes before equation 61:  ``Let $H$ be a function of the $g^{\mu\nu}$ and  their first derivatives $\frac{\partial g^{\mu\nu}}{\partial x_{\sigma} }$\ldots.'' 
This statement already guarantees energy-momentum conservation for gravity in the absence of matter, because it implements rigid space-time translation invariance for the vacuum gravitational field equations.  But Einstein has a great deal of work yet to do that  he seems to think is relevant to arriving at the conservation laws.  In particular, he pays a great deal of attention to rigid linear coordinate transformations.  
``Now we shall assume $H$ to be invariant under linear transformations, i.e., $\Delta H$ shall vanish with $\frac{ \partial^2 \Delta x_{\mu} }{\partial x_{\alpha} x_{\sigma} }$ does.''  \cite{EinsteinFormal}  
This statement is ambiguous:  clearly one could include rigid translations within this notion of linearity.  But that point is not made.  In arriving at an identity supposedly following from strictly linear transformations, invariance under translations gets smuggled in at least once.  In deriving equation 76a  with $S^{\nu}_{\sigma}$ from matter energy balance and gravity's field equations, he sets $$\frac{1}{2} g^{\mu\nu},_{\sigma} \frac{ \partial H\sqrt{-g} }{\partial g^{\mu\nu} } + \frac{1}{2} g^{\mu\nu},_{\sigma\alpha} \frac{ \partial H\sqrt{-g} }{\partial g^{\mu\nu},_{\alpha} }$$ equal to $$\frac{1}{2} (\sqrt{-g}H),_{\sigma}.$$

What if we allowed ourselves strictly linear but not affine symmetry of the Lagrangian density, such as by including a term $ A \sqrt{-g} g_{\mu\nu} x^{\mu} x^{\nu}$ ($A$  a constant)?  This term is a weight $1$ scalar density under linear (but not affine!) coordinate transformations and so transforms properly to leave the action invariant under linear (but not affine) coordinate transformations. (If one worries that this term becomes large at spatial or temporal infinity, one could use something like $A \sqrt{-g} sinc^2 (g_{\mu\nu} x^{\mu} x^{\nu})$ instead; the points that I wish to make would still hold.) 
 Let  $\xi^{\mu}= m^{\mu}_{\nu} x^{\nu}$ be an infinitesimal vector corresponding to a small version of the coordinate transformation  $x^{\mu^{\prime}} = M^{\mu}_{\nu} x^{\nu}= (\delta^{\mu}_{\nu} + m^{\mu}_{\nu}+ \ldots)  x^{\nu}$; $m^{\mu}_{\nu}$ is constant and so small that quadratic terms are negligible.   The coordinates $x^{\mu}$ are a vector under rigid linear transformations, so one can treat them as a vector field for present purposes, giving them a Lie derivative, one that vanishes in fact.  That vanishing  makes sense because the Lie derivative gives a fixed coordinate variation, comparing the value of a field at the same numerical coordinate label in two different coordinate systems \cite{BergmannHandbuch}.

Let us work through the derivation, covering the same ground a bit more quickly because the Lie derivative is now familiar (to us!).  Let $\mathcal{H}$ be the Lagrangian density $H \sqrt{-g}$, but allowing a  term like $ A \sqrt{-g} g_{\mu\nu} x^{\mu} x^{\nu}$ to be included within  the Lagrangian density also.  
Then one has the identity 
\begin{eqnarray} 
0= \pounds_{\xi} \mathcal{H}- \pounds_{\xi} \mathcal{H}  = 
\frac{\partial (\xi^{\mu} \mathcal{H})}{\partial x^{\mu} }  - \frac{\partial \mathcal{H} }{\partial g^{\sigma\tau} } \pounds_{\xi} g^{\sigma\tau}  - \frac{ \partial \mathcal{H} }{ \partial g^{\sigma\tau}_{\nu} } \pounds_{\xi} g^{\sigma\tau}_{\nu} -\frac{\partial \mathcal{H} }{\partial x^{\mu}} \pounds_{\xi} x^{\mu}        \nonumber \\ =
m^{\sigma}_{\nu} \delta^{\nu}_{\sigma} \mathcal{H} + m^{\sigma}_{\nu} x^{\nu} \frac{d \mathcal{H} }{\partial x^{\sigma} } 
- \frac{\partial \mathcal{H} }{\partial g^{\sigma\tau} } ( m^{\mu}_{\nu} x^{\nu} g^{\sigma\tau}_{\mu} - 2 g^{\nu\tau} m^{\sigma}_{\nu}) -    \frac{\partial \mathcal{H} }{\partial g^{\sigma\tau}_{\nu} } ( m^{\mu}_{\nu}  g^{\sigma\tau}_{\mu}  + m^{\mu}_{\alpha} x^{\alpha} \partial_{\mu}  g^{\sigma\tau}_{\nu}  - 2 g^{\alpha\tau}_{\nu} m^{\sigma}_{\alpha}) - 0 \nonumber \\ =
m^{\sigma}_{\nu} \delta^{\nu}_{\sigma} \mathcal{H} + m^{\sigma}_{\nu} x^{\nu} \left(  \frac{\partial  \mathcal{H} }{\partial  g^{\alpha\beta}} g^{\alpha\beta}_{\sigma}   + \frac{\partial \mathcal{H} }{\partial g^{\alpha\beta}_{\rho} }  \partial_{\sigma} g^{\alpha\beta}_{\rho}
+ \frac{\partial \mathcal{H} }{\partial x^{\sigma} } \right) 
- \frac{\partial \mathcal{H} }{\partial g^{\sigma\tau} } ( m^{\mu}_{\nu} x^{\nu} g^{\sigma\tau}_{\mu} - 2 g^{\nu\tau} m^{\sigma}_{\nu})  \nonumber \\  -    \frac{\partial \mathcal{H} }{\partial g^{\sigma\tau}_{\nu} } ( m^{\mu}_{\nu}  g^{\sigma\tau}_{\mu}  + m^{\mu}_{\alpha} x^{\alpha} \partial_{\mu}  g^{\sigma\tau}_{\nu}  - 2 g^{\alpha\tau}_{\nu} m^{\sigma}_{\alpha})  \nonumber \\ =
m^{\sigma}_{\nu} \delta^{\nu}_{\sigma} \mathcal{H} + m^{\sigma}_{\nu} x^{\nu}   \frac{\partial \mathcal{H} }{\partial x^{\sigma} }  
- \frac{\partial \mathcal{H} }{\partial g^{\sigma\tau} } ( - 2 g^{\nu\tau} m^{\sigma}_{\nu})  \nonumber \\  -    \frac{\partial \mathcal{H} }{\partial g^{\sigma\tau}_{\nu} } ( m^{\mu}_{\nu}  g^{\sigma\tau}_{\mu}    - 2 g^{\alpha\tau}_{\nu} m^{\sigma}_{\alpha})   \nonumber \\ =
m^{\sigma}_{\nu} \delta^{\nu}_{\sigma} \mathcal{H} + m^{\sigma}_{\nu} x^{\nu}   \frac{\partial \mathcal{H} }{\partial x^{\sigma} }  
+m^{\sigma}_{\nu} 2 \frac{\partial \mathcal{H} }{\partial g^{\sigma\tau} }  g^{\nu\tau}    -   m^{\sigma}_{\nu}  \frac{\partial \mathcal{H} }{\partial g^{\mu\tau}_{\nu} }   g^{\mu\tau}_{\sigma}   +  m^{\sigma}_{\nu} 2 g^{\nu\tau}_{\mu} \frac{\partial \mathcal{H} }{\partial g^{\sigma\tau}_{\mu} }   \nonumber \\ 
=2 m^{\sigma}_{\nu}  S^{\nu}_{\sigma} +  m^{\sigma}_{\nu} x^{\nu}   \frac{\partial \mathcal{H} }{\partial x^{\sigma} } =0.
\end{eqnarray} 
From the arbitrariness of $m^{\sigma}_{\nu}$ one infers that 
$$ 2 S^{\nu}_{\sigma} +   x^{\nu}   \frac{\partial \mathcal{H} }{\partial x^{\sigma} } =0.$$ 
 One easily sees that the linearly but not affinely invariant term  $ A \sqrt{-g} g_{\mu\nu} x^{\mu} x^{\nu}$ satisfies this extended identity, but not Einstein's identity $ 2 S^{\nu}_{\sigma} =0.$  Thus Einstein's identity follows not from strictly linear invariance, but from affine invariance.  But the translation invariance is what does the real work of insuring conservation laws.

Rearranging a bit, one can find the  conserved current associated with Noether's first theorem.  One has the identity
$$0= \pounds_{\xi} \mathcal{H}- \pounds_{\xi} \mathcal{H}  = \partial_{\mu} (\xi^{\mu} \mathcal{H} - \frac{\partial \mathcal{H} }{\partial g^{\sigma\tau}_{\mu} } \pounds_{\xi} g^{\sigma\tau}) - \frac{d\mathcal{H} }{dg^{\sigma\tau} } \pounds_{\xi} g^{\sigma\tau} =0.$$
Substituting the explicit form of the vector field $\xi^{\mu} = m^{\mu}_{\nu} x^{\nu},$ one obtains 
$$ \partial_{\mu} (m^{\alpha}_{\nu} [ \delta^{\mu}_{\alpha} x^{\nu} \mathcal{H} - \frac{\partial \mathcal{H} }{\partial g^{\sigma\tau}_{\mu}} x^{\nu} \partial_{\alpha} g^{\sigma\tau} + 2 \frac{\partial \mathcal{H} }{\partial g^{\sigma\tau}_{\mu}} g^{\sigma\nu} \delta^{\tau}_{\alpha} ]) - \frac{d\mathcal{H} }{d g^{\sigma\tau} } m^{\alpha}_{\nu} (x^{\nu} \partial_{\alpha} g^{\sigma\tau} -  2 g^{\sigma\nu} \delta^{\tau}_{\alpha})=0.$$
 Einstein was indeed near an application (in advance) of Noether's first theorem to his linear symmetry.  This might well have been a novelty.  
Unfortunately this conservation law does not entail the conservation of energy-momentum (certainly  not with what one hoped would be the energy-momentum). To entail energy-momentum conservation, the linear symmetry would need to exclude the term $ A \sqrt{-g} g_{\mu\nu} x^{\mu} x^{\nu}$.  But the exclusion of such terms implies/requires translation invariance. (Einstein quietly smuggled translation symmetry in by allowing only dependence on the metric and its derivative.) Once translation invariance appears, energy-momentum conservation is assured.  But then the linear symmetry is completely irrelevant, an idle wheel as far as energy-momentum conservation is concerned.  Hence Einstein applied something like Noether's first theorem in advance of Noether, but to the wrong symmetry for achieving his goal of conservation laws. It is perhaps noteworthy that the linear coordinate symmetry includes rotations and boosts and so implies the conservation of angular momentum and center-of-mass motion.

Suppose that, not noticing all this, we attempt to form the canonical energy-momentum tensor for gravity, using for a Lagrangian density  including a term like  $ A \sqrt{-g} g_{\mu\nu} x^{\mu} x^{\nu}$.  It is useful to review the derivation of the canonical tensor in the presence of explicit space-time coordinate dependence, $\mathcal{L}(g, \partial g, x)$. (Because I am not paralleling Einstein's derivation, I will feel free to use $\mathcal{L}$ as the symbol for the Lagrangian density.) The total derivative $d_{\mu}$ takes into account both implicit dependence on location \emph{via} the metric $g$ and its partial derivative, and explicit coordinate dependence. (Indices are suppressed. One can take $g$ to stand for gravity and matter if matter is included.)
 One has 
\begin{eqnarray} 
0  = - d_{\mu} \mathcal{L} + d_{\mu} \mathcal{L}= \nonumber  \\  
-\delta^{\nu}_{\mu} d_{\nu} \mathcal{L} + \frac{\partial \mathcal{L}}{\partial g} \partial_{\mu}g  + \frac{\partial \mathcal{L}}{\partial (\partial_{\nu} g)} \partial_{\mu} \partial_{\nu} g  + \frac{\partial \mathcal{L}}{\partial x^{\mu}} =  \nonumber  \\
 -\delta^{\nu}_{\mu} d_{\nu} \mathcal{L} + \frac{\partial \mathcal{L}}{\partial g} \partial_{\mu}g  + d_{\nu} \left( \frac{\partial \mathcal{L}}{\partial (\partial_{\nu} g)} \partial_{\mu}  g \right)    - d_{\nu} \left(\frac{\partial \mathcal{L}}{\partial (\partial_{\nu} g)} \right) \partial_{\mu}  g       + \frac{\partial \mathcal{L}}{\partial x^{\mu}} =  \nonumber \\ 
  d_{\nu} \left( -\delta^{\nu}_{\mu} \mathcal{L} +  \frac{\partial \mathcal{L}}{\partial (\partial_{\nu} g)} \partial_{\mu}  g \right)   + \left(\frac{\partial \mathcal{L}}{\partial g}  - d_{\nu}\frac{\partial \mathcal{L}}{\partial (\partial_{\nu} g)} \right) \partial_{\mu}  g     + \frac{\partial \mathcal{L}}{\partial x^{\mu}}   =  \nonumber \\ 
  d_{\nu} \left(   \frac{\partial \mathcal{L}}{\partial (\partial_{\nu} g)} \partial_{\mu}  g   -\delta^{\nu}_{\mu} \mathcal{L}  \right)     + \frac{d \mathcal{L}}{d g}  \partial_{\mu}  g  + \frac{\partial \mathcal{L}}{\partial x^{\mu}}  =0. \end{eqnarray}
The Euler-Lagrange equations $\frac{d \mathcal{L}}{d g}=0 $ contain the total derivative $d$ in order to get rid of boundary terms.  Energy-momentum conservation is also with respect to the total derivative  in order to try to achieve a temporally constant spatial integral.  Thus one sees that, using the field equations, the conservation of energy-momentum fails insofar as  $ \partial_{\mu} \mathcal{L} \neq  0$ (non-uniformity of nature, failure of space-time translation invariance),  as expected.  
 One could observe this failure even more explicitly, if desired, by adding to a reasonable Lagrangian density the term  
\begin{equation} 
  A \sqrt{-g} g_{\mu\nu} x^{\mu} x^{\nu}. 
\end{equation} 
This term is suitably covariant under rigid strictly linear coordinate transformations, giving an invariant contribution to the action, but it violates space-time translation invariance.  Thus one sees that strictly linear transformations are quite irrelevant, while translations do the work of ensuring conservation.  
Thus Einstein has used space-time translation invariance of $H\sqrt{-g}$ crucially while giving it only a nod by leaving $x^{\mu}$ out of the list on which $H\sqrt{-g}$ depends.  By contrast he used  the wrong symmetry, rigid strictly linear coordinate transformations, with fanfare on his way to inferring conservation laws. Far from anticipating a special case of Noether's first theorem (one that was already in the literature of his day and that had simpler analogs that were very old), Einstein had a surprisingly poor grasp of the connection between symmetries and conservation laws at this stage.  Thus he used an elaborate, quite unnecessary and misleading alternative to the simple route.  (Luckily for Einstein's definitive theory, the linearly covariant term $A \sqrt{-g} g_{\mu\nu} x^{\mu} x^{\nu}$ is not generally covariant.  General covariance also includes translation invariance, moving down to zeroth order as well as up to second and higher powers, if one represents `general' coordinate transformations using a Taylor series, as is sometimes useful \cite{OgievetskyLNC}.)

Einstein's same expression $S^{\nu}_{\sigma}$ (now multiplied by $2$) reappears in the mature 1916 treatment of the conservation laws \cite[equation 14]{EinsteinHamiltonGerman}. Hence some of the above discussion still applies to Einstein's understanding of (what we take to be) his definitive theory. 
That work concludes:
\begin{quote} 
It is to be emphasized that the (generally covariant) laws of conservation (21) and (22) are deduced from the field equations (7) of gravitation, in combination with the postulate of general covariance (relativity) \emph{alone}, without using the field equations (8) for material phenomena. \cite{EinsteinHamiltonGerman} 
\end{quote} 
Equation (21) is the true but pseudotensorial conservation law $$ \frac{\partial}{\partial x^{\nu} }(\mathfrak{T}^{\nu}_{\sigma} + \mathfrak{t}^{\nu}_{\sigma}),$$ while (22) is the material energy-momentum balance law.  
That Einstein didn't need to use general covariance to get the true conservation law if only he had used the matter field equations remains true.
One doesn't need any relativity at all, but only the $4$-parameter translation group, the uniformity of nature \cite[chapter 12]{Goldstein}.   On the other hand, for the mature theory, the gravitational field equations do in fact entail conservation by themselves \cite{EinsteinHamiltonGerman} (and even \emph{vice versa}! \cite{Noether,Anderson,EnergyGravity}).  One might find such a result very satisfying if one has it.  But one should not insist on it in advance because there is little antecedent reason to think that it is possible.  Neither is there any  evident reason to consider it necessary.  To think otherwise is akin to planning to support oneself by winning the lottery when one is able to work.  %

There are also the restrictions $B_{\mu}= 0$, Einstein's view of which has occasioned some discussion.  
  A clearer view of the logic will appear below in the discussion of ``\emph{Entwurf} electromagnetism,'' a theory of electromagnetism (not ever seriously proposed by anyone, to my knowledge) that is analogous to the \emph{Entwurf} gravitational theory.


\section{Conservation Laws: A Sketch of History}

Should we be surprised by Einstein's apparent lack of awareness of the connection between symmetries and conservation laws?  Gorelik has remarked (and I take the liberty to repeat a sentence from above):
\begin{quote} There is no indication in Einstein's works that he was aware of the
connection between symmetry and conservation principles. (He discussed
the importation into GR, not of all ten conservation laws of SR, but only of
those pertaining to energy and momentum). This is not surprising. It was
only in 1911 that Gustav Herglotz established, for the first time, a link
between ten conservation laws and ten symmetries of the Poincar\'{e} group
in the context of the mechanics of continuous media in SR (Vizgin 1972),
which was rather remote from Einstein's domain of interest. And Noether's
work in which the symmetry-conservation link was elaborated in a general form, appeared only in 1918, when Einstein had already worked out his
pseudotensor solution. \cite{GorelikConservation} \end{quote} 
I am glad to have Gorelik's agreement in the first sentence or two.  
If Einstein could only have known the basic point about the link between time- and space-translation symmetries and the conservation of energy and momentum (respectively) by encountering one recent paper (Herglotz's) on relativistic continuum mechanics,  or by reading work from the future (Noether's), then I would share Gorelik's lack of surprise as well.  However, neither the Poincar\'{e} group nor continua (whether mechanical or fields) are essential conceptual ingredients.  The essential ingredients were in fact much older, on the order of a century or more, being present already in nonrelativistic particle mechanics.  Hence if we are not surprised by Einstein there, there must be some deeper reason, perhaps such as a difference between mathematics and physics.

It has long been known that energy and momentum conservation, respectively, hold due to $\partial(laws)/\partial t=0,$  $\partial(laws)/\partial x^i= 0.$  
Historical sketches for energy, momentum, and angular momentum have appeared 
 \cite{LagrangeMecaniqueAnalytiqueNouvelle} (translated as  \cite[pp. 180-183]{LagrangeEnglish}) \cite{HamiltonConservation,Jacobi,KuhnEnergy,WignerVanDamInvariance,Elkana,KastrupNoetherKleinLie} \cite[pp. 59, 60, 62]{Whittaker}  \cite[p. 34]{KosmannSchwarzbachNoether}  including Newton, Huygens, Bernoullis, Lagrange, Hamilton and Jacobi. Kastrup's work is particularly valuable. %
The conservation of momentum 
\begin{quote} has been evolved gradually from the observation of Newton, \emph{Principia}, Book I. introd. to Sect. XI., that if any number of bodies are acted on only by their mutual attractions, their common centre of gravity will either be at rest, or move uniformly in a straight line. \ldots \cite[p. 59]{Whittaker}
\end{quote} 
 Likewise angular momentum appears in Kepler and Newton (p. 60), while energy appears in Huygens, Newton, the Bernoullis, and Lagrange (p. 62).
Here is Lagrange (until the 1990s unavailable in English):
 \begin{quote} 14. Une int\'{e}gration qui a toujours lieu lorsque les forces sont des fonctions de distances, et que les fonctions $T$, $V$, $L$, $M$, etc., ne contiennent point la variable finie $t$, est celle qui donne le principe de la conservation des forces vives.'' \cite[p. 318]{LagrangeMecaniqueAnalytiqueNouvelle} \cite[p. 34]{KosmannSchwarzbachNoether}. \end{quote}  
 With some assumptions, ``l'int\'{e}gral trouv\'{e}e sera simplement $T+V = const.$; laquelle contient le principe de la conservation des forces vives (sect. III, art. 34).'' \cite[p. 319]{LagrangeMecaniqueAnalytiqueNouvelle} 
The translation of Lagrange  renders this passage as:
\begin{quote} 14. An integration which is always possible when the forces are functions of distance and when the functions $T$, $V$, $L,$, $M$, etc. do not contain the finite variable $t$, is the one which will result in the principle of the {\bf Conservation des Forces Vives}.'' \cite[p. 233]{LagrangeEnglish} \end{quote} 
Kosmann-Schwarzbach is prepared to say ``kinetic energy'' rather than the untranslated ``Forces Vives''; the reminder of the \emph{vis viva} controversy of the 18th century is a signal not to demand a modern vocabulary.  
 ``Therefore, the derived integral will simply be $T+V=$ constant which expresses the principle of the {\bf Conservation des Forces Vives} (SECTION III, Article 34). \cite[p. 234]{LagrangeEnglish} Hamilton related the conservation of energy (to use the modern term), momentum and angular momentum to symmetries of his function \cite{HamiltonConservation}. 
During the 1830s-40s, Jacobi  discussed the  conservation of living force as holding if and only if there was no $t$-dependence \cite{HamiltonJacobiHistory,Jacobi}.

But it took a while longer for the mature understanding of conservation laws \emph{as applied to real physical systems including electromagnetism} and conservation as connected with translation invariance of a Lagrangian to become clear, because it was not so clear than as now that fundamental physics could be described in terms of a Lagrangian or Hamiltonian. Electromagnetism posed a challenge for a time \cite{HelmholtzKantConservation,ClausiusElectrodynamic,RoseveareMercury}.  
Subsumption of the fundamental physical laws (or rather their classical antecedents!) to the principle of least action clearly postdated Lagrange.  But it remains clear that, \emph{if} a variational principle is accepted, then one ought to expect a Lagrange-like result to hold in classical field theory.  Thus one might expect that adopting a variational principle would imply  inferring conservation laws from rigid translation symmetries.

Houtappel, van Dam and Wigner are useful for a discussion of less ancient sources: 
 \begin{quote} Dr. E. Guth kindly acquainted us with his study of the
history of the connection between conservation laws and the
invariance of the Lagrangian. Apparently, the first one to notice
the connection (in 1842) was C. G. J. Jacobi\ldots
who derived the conservation laws for linear and angular momentum from the
Euclidean invariance of the Lagrangian. \ldots [Re-inventors G\"{o}tt in 1897, Hamel in 1904.]
The first complete discussion of the derivation of the ten integrals
of motion (corresponding to the ten infinitesimal elements of
the inhomogeneous Lorentz group) was given by G. Herglotz
[Ann. Physik {\bf 36}, 493 (1911)].  F. Klein called attention to
Herglotz' work and encouraged F. Engel\ldots, G\"{o}tt\ldots, 
E. Noether\ldots; 
and E. Bessel-Hagen\ldots. 
 \cite[p. 606]{WignerVanDamInvariance}  \end{quote}

What did Einstein know and when did he know it?
The canonical energy-momentum tensor appeared in a mature form in Herglotz for relativistic continua  \cite{HerglotzConservationPoincare}, Mie for electrodynamics \cite{Mie1,Mie3} (who worried about asymmetry), and Born for classical field theory in general.  Born seems to have invented classical field theory  by finding the generality of which Herglotz's continua and Mie's electromagnetism were special cases \cite{BornEnergyMieHerglotz}. Born's paper was submitted by Hilbert on 20 December 1913.  
Born's understanding of the connection between rigid translation symmetries and conservation laws was crystal-clear: 
\begin{quote} 
The  assumption of Mie just emphasized, that the function $\Phi$ [the Lagrangian density]  is independent of
$x,$ $y,$ $z,$ $t,$ is also the real mathematical reason for the validity of the
momentum-energy-law.
\ldots  We assert that for these
differential equations, a law, analogous to the energy law (3$^{\prime}$) of Lagrangian
mechanics, is always valid as soon as one of the 4 coordinates $x_{\alpha}$ does not appear
explicitly in $\Phi$. \end{quote} 
  Thus there is a clear sense in which the canonical energy-momentum tensor should have been available to Einstein, or at least readily invented by him due to analogs in mechanics, during much of this period.  That Einstein was not given to long bibliographies is evident, but this weakness in scholarship did not serve him well in the context of conservation laws.  Herglotz's work would eventually be used to clarify the right side of Einstein's equations  \cite{NordstromGREnergy}. 
Born's criterion for conservation laws is satisfied by both the \emph{Entwurf} theory and the definitive theory of 1915-16.

One point that should almost go without saying is that Born's derivation of the conservation laws involves Euler-Lagrange equations for \emph{all} the fields.  That point is worth emphasizing, however, due to its contrast with Einstein's failure most of the time to use a matter Lagrangian and have Euler-Lagrange equations for matter.  If Einstein had used a matter Lagrangian as well as a gravitational Lagrangian, he would have arrived at conservation laws for energy-momentum readily due to rigid translation invariance, just as Born did, with no diversion through a linear coordinate symmetry and the detailed form of the gravitational field equations; still less would general covariance have been necessary as seems to be the case in (\cite{EinsteinHamiltonGerman}).  The fact that Einstein appealed to general covariance and avoided using the matter Euler-Lagrange equations in 1916 shows that (\emph{pace} Ohanian  \cite{OhanianEnergyNonminimal}) Einstein did not imitate Born; presumably Einstein  failed to grasp the significance of Born's work if he knew of it. (Born doesn't seem to have noticed either.) Did Einstein ever understand the relationship between conservation laws and translation symmetries?


For convenience, here is a (hardly novel, \emph{e.g.} \cite[chapter 12]{Goldstein}) derivation of the energy-momentum conservation laws for a system of gravity $g$ and matter $u$ coupled together, parallel to the  Lagrangian particle derivation of conservation of energy from  $\frac{\partial (laws)}{\partial t} =0.$
Space-time rigid translation invariance is assumed, but no higher symmetry, not even Poincar\'{e} invariance, much less linear or affine invariance, is used for the Lagrangian density  $\mathcal{L}(g, \partial g, u, \partial u, \cancel{x^{\mu}})$. $
g$ is the metric-gravitational field $g_{\mu\nu}$ (or its inverse or the like), while matter $u$ is any matter admitting variational principle. Indices of $g_{\mu\nu}$ and (as needed) $u$ are suppressed. A comma indicates partial differentiation, and leaves less clutter than sole use of $\partial$.  Much of the point is to make memorable a point that is otherwise less vividly conveyed by a summation over fields.  
\begin{eqnarray} 
0= - \partial_{\mu} \mathcal{L} + \partial_{\mu} \mathcal{L}=
 -\delta^{\nu}_{\mu} \partial_{\nu} \mathcal{L} + \frac{\partial \mathcal{L}}{\partial g} g,_{\mu}  + \frac{\partial \mathcal{L}}{\partial g,_{\nu} } g,_{\mu\nu}  + \frac{\partial \mathcal{L}}{\partial u} u,_{\mu}  + \frac{\partial \mathcal{L}}{\partial u,_{\nu} } u,_{\mu\nu}  = \nonumber \\
 -\delta^{\nu}_{\mu} \partial_{\nu} \mathcal{L} + \frac{\partial \mathcal{L}}{\partial g} g,_{\mu}  + \partial_{\nu} \left(\frac{\partial \mathcal{L}}{\partial g,_{\nu} } g,_{\mu} \right)    - \partial_{\nu} \left(\frac{\partial \mathcal{L}}{\partial g,_{\nu} } \right) g,_{\mu} + \frac{\partial \mathcal{L}}{\partial u} u,_{\mu}  + \partial_{\nu} \left(\frac{\partial \mathcal{L}}{\partial u,_{\nu} } u,_{\mu} \right) -  \partial_{\nu} \left( \frac{\partial \mathcal{L}}{\partial u,_{\nu} } \right)  u,_{\mu} = \nonumber \\
  \partial_{\nu} \left( -\delta^{\nu}_{\mu} \mathcal{L} +  \frac{\partial \mathcal{L}}{\partial g,_{\nu} } g,_{\mu}    + \frac{\partial \mathcal{L}}{\partial u,_{\nu}} u,_{\mu} \right)   + \left( \frac{\partial \mathcal{L}}{\partial g}  - \partial_{\nu}\frac{\partial \mathcal{L}}{\partial g,_{\nu} } \right) g,_{\mu}      + \left( \frac{\partial \mathcal{L}}{\partial u}  - \partial_{\nu}\frac{\partial \mathcal{L}}{\partial u,_{\nu} } \right) u,_{\mu}  = \nonumber \\ 
  \partial_{\nu} \left(   \frac{\partial \mathcal{L}}{\partial g,_{\nu} } g,_{\mu} + \frac{\partial \mathcal{L}}{\partial u,_{\nu} } u,_{\mu}   -\delta^{\nu}_{\mu} \mathcal{L}  \right)     + \frac{d \mathcal{L}}{d g} g,_{\mu}    + \frac{d \mathcal{L}}{d u} u,_{\mu} = 0.
\end{eqnarray} 
Here  $\frac{d \mathcal{L}}{d g}=0 $ are the  gravity field equations, while $ \frac{d \mathcal{L}}{d u} =0$ are the material field equations, which Einstein neglected. 
 So the conservation of material + gravity energy-momentum follows from $\frac{d \mathcal{L}}{d g_{\mu\nu}}=0 $ and $ \frac{d \mathcal{L}}{d u} =0,  $ but not from  $\frac{d \mathcal{L}}{d g_{\mu\nu}}=0. $ 
One would no more expect conservation to follow from just the gravitational field equations when matter is present, than one would expect conservation for two interacting oscillators using the equations of motion of just one of them.\begin{footnote}{Remarkably, for the `final' field equations of General Relativity, one actually does get conservation laws from the gravitational equations alone \cite{Noether,Anderson,EnergyGravity}, which is possible only because the theory has constraints that tie the gravitational energy-momentum to be equal and opposite the matter energy-momentum apart from terms with automatically vanishing divergence, a very special property.  But there was no reason to expect such a result.  Clearly Einstein and Grossmann did not expect it:  they thought that conservation failed instead, as appeared in the epigraph  \cite[p. 160]{EinsteinBesso}.  These issues are related to the case of electric charge \cite{BradingNoetherWeyl}, but complicated by the fact that gravity has energy but electromagnetism does not have charge.} \end{footnote} 
 There is no need for \emph{details} of gravity's or matter's field equations. One does not need the sophistication of Noether's first theorem \cite{Noether}.
One does not need relativistic or Galilean boost invariance; one does not even need rotational invariance unless one also demands the conservation of angular momentum.  One does not need tensor calculus.  One needs only to assume that everything with energy-momentum has Euler-Lagrange equations and that the spatio-temporal uniformity of nature holds.

Many quantum field theory textbooks begin with a treatment of classical Lagrangian field theory, wherein one sees how rigid translation invariance leads to energy-momentum conservation, among other things.  This result is so well known now that it is worth a moment's pause to ponder why its relevance to the history of General Relativity has not seemed evident.  The reason, presumably, is the idea that quantum field theory books' discussions of Noether's (first) theorem apply only to field theories that live in a certain kind of geometry, Minkowski space-time, a geometry with many symmetries; perhaps one imagines an infinite perfect grooved crystalline block underwriting certain metrical properties (a mental image, if present, which might be appropriately counteracted by calling Minkowski space-time a ``glorious nonentity''  \cite{BrownPooleyNonentity}).  By contrast General Relativity is a field theory involving a very different, variably curved space-time with (typically) no symmetries (no Killing vector fields).  Even the \emph{Entwurf} theory, with a larger-than-Poincar\'{e} symmetry group, does not live in Minkowski space-time.  The key mistake lies in thinking that the applicability of Noether's first theorem is restricted to theories with   certain geometries.  In fact Noether's theorems do not care which entities we interpret as ``geometry.''  What matters is the Lagrangian density, which transformations leave it invariant (or invariant up to a divergence, making a generalization often attributed to Bessel-Hagen but already considered in part by Noether \cite[footnote on pp. 249, 250 of the original, endnote 20 of the translation]{Noether} \cite[p. 92 endnote 20 of the translation]{KosmannSchwarzbachNoether} \cite{BesselHagen}), and which fields have Euler-Lagrange equations.  Every field (`geometrical' or otherwise) is a \emph{threat} to conservation laws \cite{TrautmanUspekhi,BradingBrown}.  To avoid this threat, it is sufficient (and more or less necessary) for each field to have either symmetries (such as not seeing any explicit dependence on space-time coordinates, to start with the basics) or Euler-Lagrange equations.  In `Special Relativity' the metric has symmetries, while everything else has Euler-Lagrange equations; in General Relativity the metric has Euler-Lagrange equations, hence in that respect acting more like matter than does the flat metric of special relativity.  The symmetry group of the Lagrangian of GR + matter (or, for that matter, of \emph{Entwurf} + matter) is larger than the symmetry group of that same matter in Special Relativity, so Noether's first theorem still applies and still yields conserved quantities \cite{Noether,TrautmanConserve,SchutzSorkin}.\footnote{One can complain that the Noether-derived pseudo-tensorial conserved quantities in General Relativity have peculiar properties.  Indeed they do, but perhaps this fact is partly a feature rather than just a bug, because at least the cardinalities of the Lagrangian symmetries and of the conservation laws thereby match \cite{EnergyGravity}. In any case it is irrelevant to theories like \emph{Entwurf}, for which Noether's first theorem applies but the second does not.}  The more the symmetries of the Lagrangian (\emph{not} of the `geometry', whatever that means), the more conserved currents there are.  Thus there is \emph{no work required} to ensure conservation laws in a gravitational field theory derivable from a Lagrangian density that satisfies the spatio-temporal uniformity of nature.  One of Einstein's heuristics, seeking energy conservation in the correct form    
$$ \frac{\partial}{\partial x^{\mu} } (\sqrt{-g}T^{\mu}_{\nu} + \sqrt{-g}t^{\mu}_{\nu})=0$$ \cite{JanssenRenn},
\emph{could not possibly fail} for any theory derivable from an action principle concerning its gravitational \emph{and material} field equations, unless it violated the uniformity of nature.  Brading's application of Noether's theorems to General Relativity points one in the correct direction \cite{BradingConserve}; some of the points also apply already to the years 1913-15 during the \emph{Entwurf} phase.  


What then happens when one introduces additional symmetries, whether rotations (standard even nonrelativistically), boosts, or general linear transformations?  
As should be clear at least to us \cite{HerglotzConservationPoincare,Noether} (though evidently it was not to Einstein), the tie between symmetries and conservation laws implies that if more symmetries are introduced, then there are more conservation laws, not fewer.  Adding rotational symmetry relates coordinate systems at different spatial orientations and yields the conservation of angular momentum.  Adding Lorentz boost symmetry of Special Relativity gives a result involving center of mass motion \cite{HillNoether} and relates conservation in different frames.  Adding linear coordinate symmetry won't falsify the above; then one has the affine group.  No matter what further symmetries are added, energy-momentum conservation will continue to hold iff rigid translation symmetries exist and because of those symmetries.  Those conservation laws will fail if and only if physical laws vary with place or time, \emph{i.e.}, there is  explicit dependence of $\mathcal{L}$ on $x^{\mu}$.  Such might happen if, \emph{e.g.}, God parted the Red Sea for the Moses and the  Israelites, or an immaterial soul acted on a brain, or a Humean failure of the uniformity of nature occurred---not the sort of example that physicists as such usually considered in the 20th century,\footnote{See (\cite{LaywineKant,vanStrienVitalismPhysics}) for 18th and 19th century discussions of the soul-body problem. I owe the latter reference to Katherine Brading.} or that would have detained Einstein the Spinozist \cite{JammerEinstein}.   %


\section{\emph{Entwurf} Lagrangian More Like Newton than Maxwell}

In pursuing the electromagnetic analogy as part of his physical strategy, Einstein paid some attention to the Lagrangian density for Maxwell's theory.   ``The Lagrangian is modeled on the Lagrangian $-\frac{1}{4} F_{\mu\nu}F^{\mu\nu}$ for the free Maxwell field.'' \cite[p. 869]{JanssenRenn}  
This analogy is especially strong if one writes it in terms of a field strength, for then it takes the plausible  form of (field strength)$^2$.  On the other hand, for some purposes it is more important how the Lagrangian density looks in terms of the potentials.  The potentials are the primitive fields in the Lagrangian and the quantities varied to find the Euler-Lagrange equations.  From that standpoint of emphasizing the potentials, does the \emph{Entwurf} Lagrangian still resemble that of Maxwell's electromagnetism?

	While Einstein's physical strategy makes use of analogies to both Newtonian gravity and Maxwell's electromagnetism, it could happen on some point that the two analogies point in different directions.  One could then ask whether he followed one or the other more closely in that respect.  Such an occasion arises with the gravitational energy-momentum tensor for the \emph{Entwurf} theory.    Consider Einstein's \emph{Entwurf} Lagrangian  $$Q= -\frac{1}{2} \sqrt{-g} g^{\mu\nu} g^{\alpha\rho} \frac{\partial g_{\rho\beta}}{\partial x^{\mu}} g^{\beta\lambda} \frac{ \partial g_{\lambda\alpha}}{\partial x^{\nu}}$$  \cite{EinsteinFormal}.
While it has an obvious tensor-related feature, namely contracted indices, it lacks another by now common tensor feature, namely, a multiplicity of terms (some of them subtracted) due to different contractions of the indices.  Instead the \emph{Entwurf} Lagrangian density has the form $(gradient)^2$, like that of a scalar theory, such as the Klein-Gordon equation less the mass term, or Nordstr\"{o}m's theory, apart from nonlinearities in the metric perturbation due to the use of $\sqrt{-g} g^{\mu\nu} g^{\alpha\rho} g^{\beta\lambda}$ rather than the gravitation-free $\sqrt{-\eta} \eta^{\mu\nu} \eta^{\alpha\rho} \eta^{\beta\lambda}$. It especially resembles a scalar theory regarding its energy-momentum tensor. For a scalar theory, the canonical energy-momentum tensor is symmetric, or close enough (recalling that one needs to use a metric to raise or lower an index), because the Lagrangian is quadratic in the field gradient and one divides and multiplies by the field gradient to get the canonical energy-momentum.

By contrast, consider a familiar and highly successful vector theory, Maxwell's electromagnetism. Historically, a recognizably modern Lagrangian density for electromagnetism in terms of potentials appeared in Schwarzschild in 1903 \cite{SchwarzschildMaxwell1,KastrupNoetherKleinLie}; others would make Lorentz invariance manifest over the next few years.  
Expressing the Lagrangian density in terms of the potential $A_{\mu}$ rather than the field strength $F_{\mu\nu}$, one obtains 
$$\mathcal{L}= -\frac{1}{2}(A_{\nu},_{\mu} A^{\nu,\mu}  - A_{\mu},_{\nu} A^{\nu,\mu}). $$
There are two terms here, a possibility that exists because of the (co)vector index on $A_{\mu}.$  Actually there are three possible terms, but 
the third one $(\partial_{\mu} A^{\mu})^2$ differs from the second $A_{\mu},_{\nu} A^{\nu,\mu} $ only by a divergence and so makes little difference.  If we retain the usual custom of ignoring $(\partial_{\mu} A^{\mu})^2$, then everything interesting hinges on the relative coefficient of  $A_{\mu},_{\nu} A^{\nu,\mu}$ to that of the default term $-\frac{1}{2} A_{\nu},_{\mu} A^{\nu,\mu}.$
In Maxwell's theory, that relative coefficient is precisely what is required to avoid a negative energy scalar (``spin-$0$'') degree of freedom that would otherwise arise from $A_0$ in the default term \cite{Wentzel}.  If there were a photon mass term $-\frac{1}{2} m^2 A^{\nu} A_{\nu},$ there would be the full remaining 3 degrees of freedom per spatial point due to the three spatial components $A_n.$ $A_0$ would be determined by the other components.   For the massless case $m=0$ there is the familiar gauge freedom that further reduces the physical degrees of freedom to two per spatial point.  As a result of having two terms in the Lagrangian density, electromagnetism has an asymmetrical canonical energy-momentum tensor.  Thus avoiding negative-energy degrees of freedom pushes one to an asymmetric canonical tensor and, for massless particles/fields (with waves traveling at speed $c$), gauge freedom.  Such considerations are now part of the basics of relativistic field theory \cite{PauliFierz,FierzPauli,Wentzel}.

Given Einstein's commitment to a symmetric material stress-energy tensor and employment of a symmetric non-canonical gravitational stress-energy tensor above, it is perhaps lucky that the \emph{Entwurf} theory's canonical stress-energy tensor is (with an index moved with $g_{\mu\nu}$) symmetric, in that he could adopt the canonical definition of stress-energy without having to reconsider his gravitational theory.  On the other hand, one could imagine that the asymmetric canonical stress-energy (pseudo)tensor of the `final' late 1915 theory could have been an obstacle to adopting that theory if Einstein's process of discovery had differed slightly from how it actually went. Should such asymmetry have been an obstacle? Why wasn't it?  This asymmetry has remedies of various sorts (Belinfante-Rosenfeld technology \cite{RosenfeldStress,Wentzel}), with different collections of virtues and vices.  For a symmetric rank 2 potential, there are 5 possible terms in the Lagrangian at quadratic order, but two of them are equal up to a divergence \cite{EinsteinFormal,Sexl}.  (If one works at the level of equations of motion rather than the Lagrangian density, there are 5 distinct terms \cite[p. 141]{Ohanian}.  It follows that such equations of motion cannot be derived from a Lagrangian density unless two of the coefficients are related.)  
Excluding negative energy spin-$1$ (vector) degrees of freedom imposes some restrictions \cite{OP,Sexl,VanN}.  The result is not quite unique even with $0$ graviton mass because unimodular general relativity, with (at the lowest order approximation) a traceless potential $\gamma^{\mu}_{\mu}=0$, is also a possibility, and in a roundabout way gives Einstein's equations up to a solution-dependent cosmological constant \cite{UnruhUGR}, or one could leave the trace unrestricted and get something like a scalar-tensor theory with an arbitrary cosmological constant \cite{SliBimGRG}.   A nonzero graviton mass should also be considered.  
Trying all possible terms is a familiar practice in particle physics, though without the benefit of the philosophical name ``eliminative induction.''

What came of the symmetric energy-momentum tensor for gravity as the definitive field equations came  into focus?  In equation 20a of (\cite{EinsteinNovember1}), Einstein writes down for gravity the canonical formula for energy-momentum, a mixed weight $1$ quantity befitting his longstanding recognition of mixed weight $1$ quantities for conservation laws.  He knows that it is a tensor only under linear transformations.  Is it symmetric (with an index moved with the metric)?  Symmetry is not manifest and seems not to be an urgent question. Substantially the same situation held later in that dramatic month \cite{EinsteinNovember4} 
and in the mature 1916 review article \cite{EinsteinFoundation} %
 (which also notes that the gravitational field equations themselves entail conservation, as shown by brute force).  And indeed the gravitational energy-momentum (pseudo)tensor for the late 1915, which inherits so much from the gravitational energy-momentum tensor of the \emph{Entwurf} theory, is indeed asymmetric.  Did Einstein have good reasons for not holding out for a symmetric quantity?   Perhaps initially he just didn't notice the asymmetry.  He wrote to de Donder on 23 July 1916 \cite{EinsteinDeDonder}:  
\begin{quote} I did not perform the somewhat tedious calculation of 
$$ t^{\nu}_{\sigma}= \frac{1}{2\kappa} \left( \frac{\partial L^*}{\partial g^{\alpha\beta}_{\nu}} g^{\alpha\beta}_{\sigma} -  \delta^{\nu}_{\sigma} L^*  \right).$$ 
\end{quote}
Not having calculated the canonical energy-momentum, he might well not have noticed  that (with an index moved) it was asymmetric. What about later on when the asymmetry became evident?  Was the desire for a symmetric stress-energy complex for gravity trumped by satisfaction with the late 1915 theory---a theory for which isolation of gravitational energy-momentum seemed less natural anyway?


\section{Logic of Conservation and `\emph{Entwurf} Electromagnetism'}

Einstein's reasoning about conservation laws has attracted attention from  historians of General Relativity. 
The topic was  confusing to the actors of the 1910s \cite[chapters  14, 15]{PaisEinstein} \cite{RoweEinsteinBianchiIdentities}, 
and significantly clarified by Noether and Klein \cite{Noether,KleinGREnergy1918,BradingBrownSymmetries}.  But the degree to which much of the discussion over the last century has been supererogatory---going beyond the call of duty by trying to get conservation laws the hard way instead of the easy way---has not been discussed much (Gorelik providing an exception noticed above \cite{GorelikConservation}).  One might hope to see Einstein's reasoning (especially during the \emph{Entwurf} period) as a special case of Noether's \emph{first} theorem, given her work's definitive clarification of such matters.  One's hope for that outcome might be strengthened by noticing how widely understood were some relevant special cases on which Einstein could draw without having to anticipate work from the future. The antiquity of knowledge that energy and momentum are conserved, respectively, due to time- and space-translation invariance was recalled above.  Especially if one has a particle physics-inspired inclination to see all relativistic field theories as similar, rather than an expectation that General Relativity is exceptional, one will be likely to ask such questions.  Of course General Relativity \emph{is} exceptional in some respects, as  the mathematics dictates.  But I have in mind questions of heuristics and presumption, for which one might or might not  be inclined to try to apply the usual rules to General Relativity depending on one's attitude against or for  GR-exceptionalism, respectively.

Though Born was part of the G\"{o}ttingen circle with Hilbert, his work was not recognized as decisive or even very relevant for understanding the origin of conservation laws in General Relativity, as one notices by Born's basically non-existent role in the story of G\"{o}ttingen's wrestling with conservation laws in General Relativity \cite{RoweGoettingenNoether,Noether,KleinGREnergy1918}. Evidently Born did not notice either. Everyone seems to have followed Einstein in turning attention to the gravitational field equations apart from material Euler-Lagrange equations.  
 Turning to contemporary authors, Smeenk and Martin have come unusually close by  appreciating  the role that Mie's and Born's works played in influencing Hilbert \cite{SmeenkMartinMieBorn}.  But the fact that Mie's and Born's canonical energy-momentum tensor already formally solved the problem for Einstein, especially  in the \emph{Entwurf} phase and in part for General Relativity, does not come up.  Presumably most authors, then and now, have been so attentive to the conspicuously novel facets of Einstein's theory, especially the gravitational field equations, that its features shared with simpler theories can be overlooked.  Works that derive conservation laws for General Relativity from translation invariance of the whole theory's (gravitational + material) Lagrangian density by Noether's theorem are not so common (\cite{GoldbergConservation,SchutzSorkin} being some examples).  Often the rush to fault  the results  by some exogenous criteria for `physical meaning'---whether the early Hilbert-Klein triviality concern that the result is a sum of a term vanishing on-shell and a term with identically vanishing divergence \cite{KleinGREnergy1917,BradingBrownSymmetries} or the now-usual criticisms of pseudotensors \cite{SchrodingerEnergy,BauerEnergy,Pauli}---inhibits noticing the fact that Noether's first theorem does apply to General Relativity, does give conserved quantities, and does give them because of `rigid' translation invariance(s!).

The logic of the conservation laws can be better understand by the introduction of a possible electromagnetic theory, which I call \emph{Entwurf} electromagnetism, because it closely parallels the Einstein-Grossmann \emph{Entwurf} theory of gravity.  In contrast to the usual electromagnetic theory, for which the Lagrangian density is the difference of two terms, the Lagrangian density for \emph{Entwurf} electromagnetism in the absence of sources is a single term $(gradient)^2$, much like a scalar theory:  
 $$\mathcal{L}= -\frac{1}{2}A_{\nu},_{\mu} A^{\nu,\mu}. $$
Like a scalar theory, this electromagnetic theory has a symmetric canonical stress-energy tensor (if one moves an index with the tacit flat metric $diag(-1,1,1,1)$).  Unlike a scalar theory, this electromagnetic theory has a negative-energy degree of freedom, thanks to Lorentz invariance, in $A_0.$  One can see the issue in the energy density
\begin{eqnarray} 
t^{0}_{0} = \frac{1}{2}(\dot{A}_m^2 + A_{m},_n^2 - \dot{A}_0^2 - A_0,_n^2),
\end{eqnarray}  
which is positive for the spatial vector potential but negative for $A_0,$ the scalar potential.

The principle of `least' action for the electromagnetic potential $A_{\mu}$ gives the electromagnetic field equations 
 $\partial_{\mu} \partial^{\mu} A^{\nu} = -4\pi J^{\nu}.$   In contrast to Maxwell's equations, 
these field equations do not require the conservation of charge, though of course they permit it.  Where then does charge conservation come from?  Is it a separate postulate, by analogy to some of Einstein's thought?  Perhaps it is in classical electrodynamics \cite{Jackson}.  But when matter is really field(s), the conservation of charge comes from the \emph{matter field equations}.  In general outline the situation is the same as for de Broglie-Proca's massive electromagnetism \cite{DeserFlux,Jackson}, though massive electromagnetism is a good theory whereas \emph{Entwurf} electromagnetism is not.  Using the electromagnetic \emph{and} material field equations, one infers for \emph{Entwurf} electromagnetism $$ \partial_{\mu} \partial^{\mu} \partial_{\nu}A^{\nu}= -4\pi \partial_{\nu} J^{\nu}=0.$$ Thus $\partial_{\nu}A^{\nu}$ satisfies the \emph{source-free} wave equation.  
 This three-derivative expression is obviously a close analog of Einstein's  restriction \cite{JanssenNemesis,EinsteinCovariance,EinsteinFormal}
$$B_{\sigma} =_{df} \frac{\partial}{\partial x^{\nu}} \frac{\partial}{\partial x^{\alpha}} \left( \sqrt{-g} g^{\alpha\beta} g_{\sigma\mu} \frac{\partial g^{\mu\nu}}{\partial x^{\beta}}\right) =0,$$  
which, however, is nonlinear.  In terms of the Lagrangian density, this condition is 
$B_{\mu}= \partial_{\sigma} \partial_{\alpha} \left( g^{\nu\alpha} \frac{ \partial H\sqrt{-g} }{\partial g^{\mu\nu},_{\sigma}  }\right)=0$ \cite{EinsteinFormal}. It says that, approximately, a divergence of the gravitational potential satisfies the source-free wave equation.  The  gravitational sector of the \emph{Entwurf} theory (that is, not invoking a matter Lagrangian and its Euler-Lagrange equations) doesn't care whether or not its source is conserved; one can feed that information in by hand or not as one wishes.

Einstein evidently thought that these conditions $B_{\mu}= 0$ had something to do with coordinates and covariance and that they needed to be fed into the formalism to get conservation laws \cite{JanssenRenn}.  It is true that, in the unfortunate and backwards way that Einstein uses these conditions, they do need to be fed in to get conservation.  But at least if one takes the \emph{Entwurf} theory to be a free-standing entity based on a variational principle (neglecting Einstein's aspirations to arrive at the \emph{Entwurf} theory from a generally covariant theory with the help of conditions $B_{\mu}= 0$ \cite{JanssenRenn}), the conditions have nothing to do with coordinates or covariance.    The conditions $B_{\mu}=0,$ if imposed to get conservation laws, amount to a \emph{freezing out of sources for some gravitational degrees of freedom by hand}. Freezing out sources (third derivatives) is better than freezing out degrees of freedom altogether as one sometimes sees \cite{Rosen73}, but it is hardly a deep insight into classical field theory.

Normatively speaking, where do these restrictions come from  in a properly sorted out field theory?  Einstein often seems to think that the restrictions hold as a consequence of energy-momentum conservation.  But where does energy-momentum conservation come from?  Once again our attention should be drawn (at least in part) to the \emph{matter field equations}.  One disanalogy between gravity and electromagnetism is that for electromagnetism, the field equations of charged matter ensure charge conservation (partly because the electromagnetic field is electrically neutral), whereas for gravity, because gravity is one of the things that has energy-momentum, both the material and gravitational field equations are needed to ensure energy-momentum conservation.  
But note especially that energy-momentum conservation does not depend on the details of the gravitational field equations, or on the postulation of coordinate   restrictions, or on the assumption of material energy-momentum balance of the specific form $\nabla_{\mu} T^{\mu\nu}=0$  that Einstein uses. (Alternate possibilities will appear below.) 
Returning to \emph{Entwurf} electromagnetism, charge conservation does not depend on the electromagnetic field equations, which are indifferent to the question.  Neither does conservation depend on the \emph{assumption} of the auxiliary condition $\partial_{\nu} \partial_{\mu} \partial^{\mu} A^{\nu}=0.$   It switches the cart and the horse.  Conservation holds due to the charged matter field equations.  The charged matter field equations and the electromagnetic field equations together imply the condition $\partial_{\nu} \partial_{\mu} \partial^{\mu} A^{\nu}=0.$  Charge conservation was never at risk (if charged matter behaves reasonably), so the detailed form of the electromagnetic field equations was not needed to ensure charge conservation.  By the same token, energy-momentum conservation did not hang by the thread of the detailed gravitational field equations as Einstein and Grossmann evidently believed (recalling the epigraph): ``but Grossmann \& I believed that the conservation laws were not satisfied'' \cite[p. 160]{EinsteinBesso}.  Conservation laws  were guaranteed as long as the theory's gravitational and matter field equations were Lagrangian-based and the uniformity of nature was assumed.

Here a comparison between the special features of Maxwell's and Einstein's theories with the more typical situation exemplified by the \emph{Entwurf} theories is useful.   Charge non-conservation would  be a \emph{miracle} in \emph{Entwurf} electromagnetism or (much more realistically) de Broglie-Proca massive electromagnetism, but it would not be a \emph{contradiction}.  The theory could consistently be coupled to a non-conserved charge current; the matter field equations then would not be playing their usual part.  One would then have $\partial_{\nu} \partial_{\mu} \partial^{\mu} A^{\nu}\neq 0.$  Likewise for energy-momentum in the \emph{Entwurf} gravitational theory: it could consistently couple to a non-conserved energy-momentum.  The matter field equations would not do their usual job, and one would have $B_{\sigma} \neq 0.$  By contrast Maxwell's theory can only be coupled to a conserved charge current.  It would be a contradiction, not a miracle, for Maxwell's equations to be true and charge conservation to be false, because   $\frac{\partial^2 }{ \partial x^{\nu} \partial x^{\mu} } F^{\mu\nu} \equiv 0$ by symmetric-antisymmetric cancellation. %
 Thus the restriction on $\partial_{\nu} A^{\nu}$ disappears in favor of gauge freedom.  Likewise with the definitive Einstein equations:  it would be a contradiction, not a miracle, for Einstein's equations to be true and energy-momentum conservation to be false.  Thus the  restriction involving $B_{\sigma}$ becomes vacuous in favor of permitting arbitrary coordinates.  Such properties of Maxwell's and Einstein's theories are fascinating, perhaps even heroic.  They are not necessary, however; hence it is supererogatory and misleading to focus attention of them.  Neither are such properties typical.  The typical situation is that if the matter and gravity laws are globally space-time translation invariant, then the gravity field equations $\frac{d \mathcal{L}}{d g_{\mu\nu} }=0$ \emph{and} the matter field equations $\frac{d \mathcal{L}}{d u}=0$ together conserve energy-momentum.  Any half-reasonable theory would suffice; conservation could fail if and only if a miracle (or a Humean non-uniformity of nature) occurred.  Einstein's frequent failure to invoke the matter field equations made it seem that gravity had to conserve energy all alone.  This overly strong demand, remarkably enough, is actually fulfilled in the definitive late 1915 field equations.  But this is hardly the result of clear understanding  of the relationship between symmetries, conservation laws and Euler-Lagrange equations. 


\section{Einstein's Daring Premises and Energy Conservation} 

The imperfections of Einstein's appeal to energy-momentum conservation can best be seen if one casts it in terms of a more formal deductive argument.  What appears is that Einstein uses very strong premises that probably only he believed in order to derive conservation, when in fact conservation followed from vastly weaker and more widely acceptable premises.  Moreover, the resources needed for that superior argument were already old and widely diffused in the literature in general outline, and available in mature form in recent literature as well.  We will need some abbreviations for premises:

${\bf P}$:  the field equations for matter and gravity are, at least, {\bf P}oincar\'{e}-invariant. 
Part of Einstein's project was to generalize Special Relativity, so he would not wish to give up the symmetries already achieved.  In particular, rigid space-time translation invariance is included.  

${\bf GM}$:  {\bf G}eneral covariance applies to the {\bf M}aterial field equations.
$GM$ is a rather strong, distinctively Einsteinian premise.  It plays a key role in Einstein's argument, but no role in the conservation of the canonical energy-momentum tensor. 
Obviously $GM$ entails the matter sector of $P$, though I haven't bothered to supply a name to the Poincar\'{e} invariance of the matter laws.  

${\bf AM}$:  there is an {\bf A}ction principle for the {\bf M}aterial field equations.  Einstein only occasionally explicitly uses this premise, but he eventually accepts it, leaving little point in avoiding it.  His initial derivation of the material energy-momentum balance equation $GM\land AM \rightarrow (\nabla_{\mu}\sqrt{-g}T^{\mu}_{\nu}=0)$ for a point mass also uses a matter action.  In fact it is true for any matter fields, though Einstein seems not to make the argument.  Instead he seems to \emph{postulate} $\nabla_{\mu}\sqrt{-g}T^{\mu}_{\nu}=0$ by analogy from the result derived from the point particle action principle. 

Such postulation is somewhat peculiar logically, because it is not easy to see why that equation would be true unless it is because of $GM \land AM.$  As  Blanchet noted, it would be straightforward to handle all the empirical support for the equivalence principle without making matter see exactly an effective curved geometry.  One could, for example, simply take a generally covariant matter Lagrangian, make a perturbative expansion of the curved metric about a flat background, and then truncate after a term or two, or 47, or 500,000,000.  Indeed this sort of strategy was viable in 1992  \cite{BlanchetNonmetric}  (and presumably still is, if the constraints are merely empirical), so it was certainly viable in the 1910s. Blanchet writes:
\begin{quote}  
A class of generically nonmetric couplings between nongravitational fields and the gravitational field is
introduced. It is shown that these couplings violate the weak equivalence principle at a level (\emph{a priori})
smaller than the current E\"{o}tv\"{o}s-type experimental limits.\ldots

Metric theories of gravity, i.e., theories embodying the
Einstein equivalence principle, have deserved a privileged
status among alternative theories for nearly eighty years.\ldots
In a metric theory, the nongravitational fields of all types
are coupled, in a universal way, to a single curved spacetime
metric $g_{\mu\nu}$ associated with gravity ($g_{\mu\nu}$ is called the
physical metric), in the local Lorentz frames of which the
standard special-relativistic laws of physics are valid [references suppressed].
 This coupling to gravity, which is referred to as the
metric coupling, consists merely in the minimal replacement,
in the Lagrangian density of special relativity
(written in generally covariant form), of the Minkowski
metric $\eta_{\mu\nu}$ by the physical metric $g_{\mu\nu}$.\ldots

The primary motivation underlying the postulates of
metric theories of gravity is an empirical one.\ldots

However, we show in this Letter that the empirical
motivation for considering only metric theories of gravity
may not be fully justified at present. We introduce a
large class of generically \emph{nonmetric} couplings to gravity,
which thus do not in general satisfy the Einstein equivalence principle nor the weak equivalence principle.\ldots 

We can understand the small magnitude of the
equivalence-principle violation as follows. First of all, the
nonmetric couplings we consider have the property of being
metric at first order in the field (in a sense made precise
below).    \cite{BlanchetNonmetric}
\end{quote} 
For such a theory, the matter action will have a generalized Bianchi identity that becomes, using the matter field equations \cite{Kraichnan,Rosen73,MassiveGravity3}, 
\begin{eqnarray} 
g_{\alpha\nu}  \nabla_{\mu} \frac{ \delta S_{mt} }{ \delta g_{\mu\nu} }   +  \eta_{\alpha\nu}  \partial_{\mu} \frac{ \delta S_{mt}  }{ \delta \eta_{\mu\nu} } |g=0.
\end{eqnarray}  
If the term involving  $\eta_{\mu\nu}$ vanished, then the first term would survive to give $\nabla_{\mu}\sqrt{-g}T^{\mu}_{\nu}=0$ using Hilbert's metric definition of the stress-energy tensor using the curved metric, as desired. This is the familiar form of the energy-momentum balance law. But nonmetric coupling of the matter, even if metrical to a high degree of approximation (so that the flat background metric is difficult to notice), would spoil Einstein's energy-momentum balance equation: $\nabla_{\mu}\sqrt{-g}T^{\mu}_{\nu} \neq0$, because the flat geometry would still be detectable (however  weakly), as in the equation above.  Thus any merely empirical motivation for $\nabla_{\mu}\sqrt{-g}T^{\mu}_{\nu}=0$ as an exact relation is tenuous. If one's only logical path to a true conservation law $ \frac{\partial}{\partial x^{\mu} } (\sqrt{-g}T^{\mu}_{\nu} + \sqrt{-g}t^{\mu}_{\nu})=0$ were \emph{via} the material energy-momentum balance law in the fully metrical form $\nabla_{\mu} \sqrt{-g}T^{\mu}_{\nu}=0$ (as Einstein seemed to think), then one would have made the true conservation law tenuous.  By contrast, the true conservation law $ \frac{\partial}{\partial x^{\mu} } (\sqrt{-g}T^{\mu}_{\nu} + \sqrt{-g}t^{\mu}_{\nu})=0$ is in fact extremely robust, holding for any (gravitationally \emph{and materially}) Lagrangian-based local field theory with at least rigid translation symmetry and \emph{a fortiori} any Lagrangian-based local relativistic (at least Poincar\'{e}-invariant) field theory.  Thus it is clear how inadvisable it is to rely on the energy-momentum balance law in the purely metrical form  $\nabla_{\mu} \frac{ \delta S_{mt} }{ \delta g_{\mu\nu} }=0$ to establish the true conservation law.  One's premise could easily be false, while the conclusion would still be true for other reasons.  

A few more premises still need to be identified and labeled.  

${\bf DG}$:  the as-yet unspecified {\bf D}etails of the {\bf G}ravitational field equations (beyond Poincar\'{e} invariance, locality, second order derivatives, \emph{etc.} that often go without saying), to be filled in so as to achieve (among other things) energy-momentum conservation.  

${\bf AG}$:  there is an {\bf A}ction principle for the {\bf G}ravitational field equations.

Einstein's argument, or argument family (taking into account the vagueness of $DG$), might be reconstructed as
$$(\nabla_{\mu} \sqrt{-g}T^{\mu}_{\nu}=0) \land P \land DG {\rightarrow}  
 \frac{\partial}{\partial x^{\mu} } (\sqrt{-g}T^{\mu}_{\nu} + \sqrt{-g}t^{\mu}_{\nu})=0.$$  
If we ask why one should accept $\nabla_{\mu} \sqrt{-g}T^{\mu}_{\nu}=0,$ recalling the hole in the aspiring commutative diagram, one might invoke $GM\land AM$:  the matter field equations are generally covariant and follow from an action principle.  Einstein's mature statement did introduce a matter Lagrangian \cite{EinsteinHamiltonGerman}, even if he emphasized that his derivation of energy-momentum conservation did not use the matter Euler-Lagrange equations.    If one reconstructs Einstein's argument along those lines, one has
$$GM\land AM \land P \land DG {\rightarrow}  
 \frac{\partial}{\partial x^{\mu} } (\sqrt{-g}T^{\mu}_{\nu} + \sqrt{-g}t^{\mu}_{\nu})=0.$$
Einstein adopted an action principle for gravity earlier and with more enthusiasm than he did for matter.  It will not be unfair to introduce premise $AG$ then: 
 $$GM\land AM \land P \land DG \land AG {\rightarrow}  
 \frac{\partial}{\partial x^{\mu} } (\sqrt{-g}T^{\mu}_{\nu} + \sqrt{-g}t^{\mu}_{\nu})=0.$$

What is peculiar about Einstein's work (partly with Grossmann), in light of the canonical tensor derivation tradition from Lagrange to Born, is that the \emph{premises} $GM$ and $DG$ \emph{are wholly irrelevant}---neither the general covariance of the matter laws nor the detailed form of the gravitational field equations is needed.  The material energy-momentum balance lemma of the form  $\nabla_{\mu} \sqrt{-g}T^{\mu}_{\nu}=0$ is also irrelevant.  They are irrelevant because it is already the case that 
$$AM \land P \land AG {\rightarrow}  
 \frac{\partial}{\partial x^{\mu} } (\sqrt{-g}T^{\mu}_{\nu} + \sqrt{-g}t^{\mu}_{\nu})=0.$$
Conservation follows from $P$, Poincar\'{e} invariance of all the laws (material and gravitational) and from the Euler-Lagrange equations for both gravity ($AG$) and matter $AM$.  Indeed one doesn't strictly need to rotations and boosts of the Poincar\'{e} group to get conservation, though they were not in doubt in Einstein's mind or in Special Relativity. It would strain out gnats and swallow camels to regard $AG$ (an action principle for gravity) as insecure and hence worth avoiding, but $GM$ (general covariance of matter laws) as a secure foundation.  It is also unclear why an action principle would be appropriate for gravity and not for matter.  Hence the canonical tensor derivation along the lines of Lagrange and Born is applicable, simple, and well motivated.  


\section{Why Did Einstein Seek What He Couldn't Lack?}

Given that energy-momentum conservation was never at risk (notwithstanding Einstein and Grossmann's beliefs and despite Einstein's efforts to use it as a restrictive principle), that the knowledge needed to realize that was both old (Lagrange \emph{et al.}) and readily available recently (Herglotz, Mie, Born), the question arises, why did Einstein try (in an incorrect way) to find what he couldn't possibly have lacked?\footnote{I thank Alex Blum for helpful discussions of this question.}  Why didn't he take conservation to be already secured by space-time translation invariance, as in Herglotz, Mie and Born, or even Lagrange, Hamilton and Jacobi?  One possibility is that he took translation invariance (the uniformity of nature) so much for granted that it didn't occur to him to draw conclusions from it.  But plenty of other people had done so.  Perhaps he didn't realize that translation invariance secured the conservation laws?  That seems like a surprising thing for a theoretical physicist working on space-time in the 1910s not to know, or at least not to suspect based on the early-mid 19th century Lagrangian particle mechanics analogs.  But Einstein is known not to have read widely in the literature \cite[p. 90]{KennefickWaves} and his bibliographies are rather brief.

It is true that variational principles were more highly regarded and better known among  mathematicians interested in mechanics than they were among physicists.  Corry has helpfully discussed Hilbert's reading list \cite{CorryHilbertGray}. Perhaps  Einstein's surprising lack of understanding into the connection between symmetries and conservation laws in a Lagrangian context is explained by his being definitely a physicist and only a late convert to mathematics.\footnote{I thank  Dennis Lehmkuhl for a discussion of this question.}  On the other hand, the connection between momentum conservation and the independence of the Lagrangian from spatial coordinates for particle mechanics---think of a free particle, a harmonic oscillator, or center-of-mass coordinates for the two-body problem---is right there on the surface:  $$ \frac{\partial L}{\partial x} -  \frac{d}{dt} \frac{\partial L}{\partial \dot{x}}=0.$$  It is difficult to imagine developing the facility to use a variational principle for a field theory with a $10$-component field, as Einstein did, without encountering or rediscovering how ignorable coordinates imply momentum conservation.  Hence the question seems to me not yet adequately resolved.

A more flattering suggestion is that Einstein wanted to avoid a matter Lagrangian density.  So often he avoids invoking one or even the matter field equations (whether or not derived from an action principle) where they would help a great deal. 
Einstein also did not love variational principles wholeheartedly.
According to Lanczos, who claimed to be of one mind with the later Einstein's ``cosmic philosophy,'' ``Einstein never cared much for `action principles','' \ldots \cite[p. 33]{LanczosEinstein}.  Einstein said as much himself to de Donder: 
\begin{quote} 
I must admit to you that, unlike most of our colleagues, I am not at all of
the opinion that every theory must be put into the form of a variation principle.
I attribute this common urge simply to the fact that everyone is used to scalars
but not to tensors. [note suppressed] If the Hamiltonian form is desired nonetheless, the following
device is best used. [Drop the second derivatives by partial integration\ldots.] 
\cite[pp. 235, 236]{EinsteinDeDonder} \cite[pp. 27,28]{KosmannSchwarzbachNoether} \end{quote}  
Einstein also wrote to Lorentz: 
\begin{quote} 
I myself am compelled to 
derive the Hamiltonian function retroactively, in order to derive the expression for
the conservation laws conveniently. [note suppressed] Nevertheless, I must admit that I actually
do not see in Hamilt. princip. anything more than a means toward reducing a
system of tensor equations to a scalar equation for which the conservation laws
are always satisfied and easily derived. \cite{EinsteinLorentz184} \end{quote}   %
But this was a day or two after he had written to Lorentz in his previous letter:
\begin{quote}
I am of the conviction that the depiction of the theory would gain much clarity
if Hamilton's formulation were taken as a point of departure, as you have done
in your fine paper published by the Amsterd[am] Acad. [note suppressed] \cite{EinsteinLorentz183} \end{quote} 
Einstein's confession of not being fully enamored of variational principles also has to be weighed against the fact that he first motivated the material energy-momentum balance law with a particle matter Lagrangian already in 1913 \cite{EinsteinEntwurfGerman}, and that he consistently used a gravitational Lagrangian once he adopted it.  Eventually variational treatments of General Relativity were produced by a number of authors \cite{Hilbert1,LorentzEnergy,EinsteinHamiltonGerman,WeylAction,NordstromGREnergy}. In 1916 Einstein emphasized that his derivation of the true conservation and material balance laws used only the gravitational field equations (to recall from above): 
\begin{quote} 
It is to be emphasized that the (generally covariant) laws of conservation (21) and (22) are deduced from the field equations (7) of gravitation, in combination with the postulate of general covariance (relativity) \emph{alone}, without using the field equations (8) for material phenomena. \cite{EinsteinHamiltonGerman} 
\end{quote} 
But he nevertheless used a matter Lagrangian density $\mathfrak{M}$ and derived material Euler-Lagrange equations (8)
$$ \frac{\partial}{\partial x^{\alpha} } \left( \frac{\partial \mathfrak{M}}{\partial q_{(\rho)\alpha}}\right) - \frac{\partial \mathfrak{M}}{\partial q_{(\rho)}} =0$$
from it!  It seems peculiar to suppose that Einstein's mature statement of the variational treatment of his theory contains assumptions that he considers basically unwarranted (leaving aside the quibbles under equation (8) that there might be higher derivatives or that the $q_{(\rho)}$ might not be independent).  If one has a matter Lagrangian, why not use it?  But then the mystery returns.

Did Einstein perhaps, by analogy to how Maxwell's equations alone entail the conservation of charge, think that the gravitational field equations alone ought to entail the conservation energy?  His physical strategy did indeed involve an electromagnetic analogy.  But Einstein's analogy to electromagnetism was not very sophisticated \cite{EinsteinEntwurfGerman} and not much like what automatically comes to mind from later writers  \cite{PauliFierz,FierzPauli,Gupta,GuptaReview,Ohanian}.  Einstein talks about electrostatics more and Maxwell's electromagnetism less than a modern person would expect.  It is indeed an interesting idea to ask that the electromagnetic or gravitational field equations alone entail their respective source.  But it is an act of theoretical supererogation, which Einstein clearly did not recognize.  Recall again the epigraphic quotation from Einstein's letter to Besso on December 10, 1915: 
\begin{quote} 
\ldots but Grossmann \& I believed that the conservation laws were not satisfied and  Newton's law did not result in first-order approximation.'' \cite[p. 160]{EinsteinBesso} 
\end{quote} 
This is not the statement that one makes if one sought a super-elegant route to conservation but had to settle for a more pedestrian route.  Rather, Einstein and Grossmann had thought that there was no other route to conservation than the one that, they concluded, had failed.  That was an underwhelming thing to think in 1915.  

Another possibility that one might consider is that Einstein wanted a symmetric energy-momentum tensor
but worried that he didn't have one.  Above his commitment to a symmetric material stress-energy tensor was exhibited. Asymmetry was of concern to Mie and Born \cite{Mie3,BornEnergyMieHerglotz}.   There is no obvious reason to value highly a symmetric material stress-energy tensor but not require a symmetric gravitational energy-momentum tensor.  But the \emph{Entwurf} theory \emph{has} a symmetric stress-energy tensor (moving the index with the curved metric)!  
I am not aware of Einstein's worrying about an asymmetric stress-energy complex from a generally covariant theory during either the first or the second phase of his weighing of such theories.  As everyone now knows, the ``Einstein pseudotensor'' (derived using the canonical recipe) is asymmetric.  But Einstein seems not to have known that during the early contemplation of a generally covariant theory.  Neither does he seem to have worried about it later; the strong 1912 commitment to symmetric stress-energy \cite{EinsteinSR1912} seems to have been overlooked or quietly waived for gravity in 1915-16.
Recall what he wrote to de Donder on 23 July 1916 \cite{EinsteinDeDonder}:  
\begin{quote} I did not perform the somewhat tedious calculation of 
$$ t^{\nu}_{\sigma}= \frac{1}{2\kappa} \left( \frac{\partial L^*}{\partial g^{\alpha\beta}_{\nu}} g^{\alpha\beta}_{\sigma} -  \delta^{\nu}_{\sigma} L^*  \right).$$ 
\end{quote}
Or did he no longer care, as more modern authors would advise?  If one doesn't believe in gravitational energy, or believes that it exists but isn't anywhere in particular (not localizable), one might not object.  (Unbelief in gravitational energy was associated with doubt about gravitational radiation, which was dispelled partly by thought experiment  \cite{Feynman,BondiNature,DeWitt57ChapelHillRickles,KennefickWaves}.) %
But Einstein continued to take gravitational energy and true conservation laws seriously during the 1910s, even when others didn't.

Einstein's reasoning regarding energy conservation, far from being prescient, was not even particularly well informed in this respect.  It fell below that of some of his contemporaries (such as Born) and predecessors (such as Lagrange), and apparently dragged 
  the G\"{o}ttingen mathematicians down to his own level by distracting attention from the matter field equations and the symmetries of the combined gravitational-material Lagrangian.  The particle physics-inspired disposition to see General Relativity as a law-abiding example of the rules of classical field theory, rather than the great exception, is sometimes fruitful in suggesting questions for the history of General Relativity.  That is perhaps all the more fitting because the \emph{Entwurf} theory has no gauge freedom, so Noether's first theorem applies but not her second.   On the other hand, Einstein's recognition that conservation laws involve a partial rather than covariant derivative and his attachment to the mixed (one index up, one down) weight $1$ form of the conservation laws even before he had the canonical tensor (to say nothing of Noether's theorems) shows him more insightful in 1913 than many of his successors in the following century.  Regarding the question of how hard one should try to make Einstein look deeply rational \cite{JanssenWhatDidEinsteinKnow}, the examples considered in this paper show Einstein capable of surprises in both directions.


\section{\emph{Entwurf}, Negative Energy and Instability?}

Having negative-energy and positive-energy degrees of freedom interact seems scary, likely to imply instability, to physicists of various times and places.  
And yet due to relativity's $-+++$ geometry, negative-energy degrees of freedom tend to crop up regularly if one is not careful \cite{FierzPauli,VanN}.
For a vector potential, if the spatial components have positive energy, then the temporal component will have negative energy unless one engineers it away, as occurs in Maxwell's theory.  Particle physicists call negative-energy degrees of freedom ``ghosts.''  Ghosts are almost always considered fatal (except in certain contexts where they are introduced as a technical tool).  What is to keep nothing from turning spontaneously into something and anti-something?  Even energy conservation, supposed to exclude perpetual motion machines of the first kind, fails to stop the bleeding.  Negative energy degrees of freedom were tacitly assumed not to exist in 19th century formulations of energy conservation, it would seem.  Lagrange considered whether positive energy was required for stability; he showed that bad things could happen if the potential were indefinite \cite{LagrangeMecaniqueAnalytiqueNouvelle}. 
 It did not occur to him to entertain negative \emph{kinetic} energy, however, something hardly conceivable apart from relativity and the de-materialization of matter into fields in the 20th century.  Lagrange showed that positive energy was stable with a separable potential  \cite{LagrangeMecaniqueAnalytiqueNouvelle}.  The separability requirement was removed by Dirichlet \cite{Dirichletstability,Morrison}. It is perhaps still not perfectly clear whether ghosts so are fatal in classical field theory \cite{BambusiGrebertTame}.  But the generally shared expectation that gravity should be quantized makes such a refuge of limited value---at least if field theory is formulated with infinite spatial volume, hence without  wavevector restrictions and therefore with  dangerous resonances.

How does the \emph{Entwurf}   theory fare if one is suspicious enough of negative energy to check for it?  
The Lagrangian density is (not worrying about an overall factor which varies from paper to paper) 
$$Q= -\frac{1}{2} \sqrt{-g} g^{\mu\nu} g^{\alpha\rho} \frac{\partial g_{\rho\beta}}{\partial x^{\mu}} g^{\beta\lambda} \frac{ \partial g_{\lambda\alpha}}{\partial x^{\nu}}$$   \cite{EinsteinFormal} \cite[p. 869]{JanssenRenn}. 
One can neglect the nonlinearity and keep only quadratic terms in the Lagrangian (hence linear terms in the field equations) using the expansion  $g_{\mu\nu}= diag(-1,1,1,1) + \sqrt{32 \pi G} \gamma_{\mu\nu}.$ Setting  $32 \pi G=1$ for convenience, one has
 $$ Q  = -\frac{1}{2} \frac{\partial \gamma_{\rho\beta}}{\partial x^{\mu}} \frac{\partial \gamma^{\rho\beta}}{\partial x^\mu}+\ldots= \frac{1}{2}(\dot{\gamma}_{00}^2 {\bf -2 \dot{\gamma}_{0i}^2} + \dot{\gamma}_{ij}^2)+\ldots:$$
\emph{Entwurf has ghosts}! The time-space components of the metric  $g_{0i}$ are 3 real degrees of freedom, and they have the wrong sign (negative energy).
Testing for positive energy became routine in particle physics from 1930s \cite{FierzPauli}.  Did Einstein or anyone seek positive energy during the 1910s?
Reading the main papers regarding the \emph{Entwurf} theory hardly shows such concern.  
 Gravitational waves make a good test, but they seem not to be studied before \cite{EinsteinApproximative}, when he is surprised by $0$-energy waves.
If he had wrestled with negative-energy waves, then he'd have been \emph{relieved} by $0$-energy waves!    Approximate wave solutions were still fresh in \cite{EinsteinEnergyWaves}\cite[pp. 251-252]{WeylSTM}\cite{EddingtonWavesEnergy}.


\section{In What Sense Was Rotation the Nemesis of  Entwurf?}

  Rotation has been described as the ``nemesis'' of the \emph{Entwurf} theory \cite{JanssenNemesis}.  As far as Einstein's self-understanding is concerned, clearly the failure of even Minkowski space-time in rotating coordinates to satisfy the field equations seemed very problematic \cite{EinsteinSommerfeld}.  
But if we set aside the context of discovery and ask ourselves why \emph{we} should reject the \emph{Entwurf} theory, or even why Einstein's contemporaries should 
reject the theory (given that they likely didn't share his opinions about the relativity of rotation), the picture changes.  Einstein had many fascinating opinions, often called Principles, some of which are borne out in his (from our perspective) definitive theory, some of which are not.  Those that apply in his final theory are at least approximately empirically adequate, which is significant, even if too weak to make them probably true without further argument \cite{UnderdeterminationPhoton}.  The problem is that the easiest way to make a plausible rival theory that also empirically approximates it arbitrarily well, adding a photon/graviton mass term, radically changes the conceptual features of the theory by destroying the gauge freedom \cite{PauliFierz,FierzPauli,FMS}.  That is at least the natural thing to think; it is rigorously true in electromagnetism even under quantization, but gravity has a long, complicated, and (since 2010) very dynamical story to which I alluded in a footnote above.  Einstein's desire to include rotating coordinate systems works in General Relativity, but it could easily have been the case (and might yet be the case) that plausible, empirically viable theories lack that feature.  Hence rotation might have been a nemesis for \emph{Entwurf} in Einstein's mind, but need not be in ours.  What \emph{really} was wrong with the \emph{Entwurf} theory is that it had ghosts.  That bug certainly would have been pointed out immediately in a more rational counterfactual history in which gravity is sorted out after special relativistic wave equations rather than before due to Einstein's premature speculative brilliance  \cite[p. 334]{OhanianEinsteinMistakes}.

How significant such a rationally reconstructed history is depends in part on what one is doing history for.  If one wants, with Ranke, just  wants only to show what actually happened, then stories about what should have happened are clearly off-target.  If one wants to exhibit the growth of genuine scientific knowledge in the spirit of justified true belief, then as Lakatos urged, one needs standards to distinguish progress from mere change  \cite{LakatosFalsification,LakatosHistory}.
The natural sciences (at least most of them) are epistemically very impressive, a fact that makes their history particularly interesting and important.  Einstein's General Relativity, in turn, is a spectacular example within the sciences.  Thus much of the interest in the history of General Relativity is rooted in the epistemic impressiveness of General Relativity.  Surely both the epistemic status of Einstein's process of discovery and the epistemic status of our degree(s) of belief in the theory are very interesting indeed.  Thus rather than leave impressive analytical tools sitting on the table unused, one should try to answer those epistemic questions as well as possible.  Particle physics provides many novel and sharp tools for that purpose.  Clearly one does not want artificial Lakatosian quasi-history with the actual contingent history in the footnotes, \emph{pace Lakatos} \cite[pp. 106, 107]{LakatosHistory}; 
as Kuhn said, such a product is ``not history at all but philosophy fabricating examples'' \cite{LakatosKuhn}. The other way around, a true history of what actually happened footnoted with remarks about what should have happened, is far more promising.  This work can be viewed as such a footnote to the history of General Relativity. This work might also be seen as an early gesture toward  Chang's vision of normatively informed history of science \cite{ChangPhlogistonWhig,ChangWater} as applied to gravity and space-time. Recall that 
 Kuhn was quite willing for historians of science to diagnose mistakes in past scientists' work; it was quite legitimate for the historian as such to discuss 
\begin{quote} \ldots  the failure of the man who creates a new theory and of his entire generation to see in that theory consequences which a later generation found there.\ldots [and] \ldots  mistakes or of what a later generation will see as having been mistakes and will accordingly feel constrained to correct.
Historical data of these sorts are all central and essential for the internal historian of science. Often they provide his most revealing clues to what
occurred. \cite{LakatosKuhn} \end{quote}
Indeed.


\section{Acknowledgements}
This work was supported by John Templeton Foundation grant \#38761. 
I thank Michel Janssen, Alex Blum, Dennis Lehmkuhl, J\"{u}rgen Renn, Harvey Brown, Jeremy Butterfield, Carl Hoefer, Gennady Gorelik, Vladimir Vizgin, Yuri Balashov, Katherine Brading and an audience member in Duesseldorf for assistance and  discussion.  All errors are my own.



\end{document}